# A Systematic Literature Review on LLM Defenses Against Prompt Injection and Jailbreaking: Expanding NIST Taxonomy

Pedro H. Barcha Correia[a,*], Ryan W. Achjian[a], Diego E. G. Caetano de Oliveira[c], Ygor Acacio Maria[a], Victor Takashi Hayashi[a], Marcos Lopes[b], Charles Christian Miers[c], Marcos A. Simplicio Jr.[a]

[a] *Universidade de São Paulo (USP), Laboratório de Arquitetura e Redes de Computadores (LARC), Brazil*
[b] *Universidade de São Paulo (USP), Departamento de Linguística, Brazil*
[c] *Universidade do Estado de Santa Catarina (UDESC), Programa de Pós-Graduação em Computação Aplicada (PPGCAP), Brazil*

**Abstract**

The rapid advancement and widespread adoption of generative artificial intelligence (GenAI) and large language models (LLMs) has been accompanied by the emergence of new security vulnerabilities and challenges, such as jailbreaking and other prompt injection attacks. These maliciously crafted inputs can exploit LLMs, causing data leaks, unauthorized actions, or compromised outputs, for instance. As both offensive and defensive prompt injection techniques evolve quickly, a structured understanding of mitigation strategies becomes increasingly important. To address that, this work presents the first systematic literature review on prompt injection mitigation strategies, comprehending 88 studies. Building upon NIST's report on adversarial machine learning, this work contributes to the field through several avenues. First, it identifies studies beyond those documented in NIST's report and other academic reviews and surveys. Second, we propose an extension to NIST taxonomy by introducing additional categories of defenses. Third, by adopting NIST's established terminology and taxonomy as a foundation, we promote consistency and enable future researchers to build upon the standardized taxonomy proposed in this work. Finally, we provide a comprehensive catalog of the reviewed prompt injection defenses, documenting their reported quantitative effectiveness across specific LLMs and attack datasets, while also indicating which solutions are open-source and model-agnostic. This catalog, together with the guidelines presented herein, aims to serve as a practical resource for researchers advancing the field of adversarial machine learning and for developers seeking to implement effective defenses in production systems.

## 1. Introduction

Large language models (LLMs) have rapidly become central to modern digital infrastructure, driving generative artificial intelligence (GenAI) applications such as search engines, code assistants, and conversational systems used daily by millions [1]. This technology has also expanded into sensitive domains through autonomous agents that perform critical actions in healthcare [2, 3] and techniques like Retrieval-Augmented Generation (RAG), which enable integration with private databases and documents [4, 2]. However, as with many disruptive technologies [5], the rushed adoption of LLMs has outpaced both regulatory frameworks and cybersecurity measures, raising ethical concerns and rendering vulnerable an increasing number of LLM-integrated systems.

Among several possible attacks against GenAI and LLMs [6], prompt injection (PI) is particularly concerning. Ranked by the Open Worldwide Application Security Project (OWASP) as the biggest risk for LLMs, PI manipulates AI models through maliciously crafted inputs [7]. According to the U.S National Institute of Standards and Technology (NIST), these attacks may seek to disrupt system operation, manipulate outputs to serve attacker goals, and extract confidential data [1]. Real-world examples include hidden prompts in academic papers that command positive reviews, biasing conversational agents that may be used by reviewers [8]; hidden injections within legal documents seeking to sway legal procedures and their outcomes [9]; and injections that can persistently leak user input in OpenAI ChatGPT [10] or other AI-enabled chatbots. Moreover, a subclass of PI known as jailbreaking [1] may be used to circumvent LLMs' safety alignment and enable misuse. This can lead widely available models like ChatGPT, Meta llama, Anthropic Claude, and Google Gemini to generate outputs related to illegal activity, hate speech, sexism, misinformation, and other harmful content [11].

To address these problems, several techniques have been developed to mitigate the impact of prompt injection on LLMs. As seen in Figure 1, the landscape is diverse, ranging from simple filters using lists of blocked terms to fine-tuning techniques in which post-training methods are employed to align model behavior with human preferences. To assist researchers and developers in making GenAI-based systems safer, this work performs a systematic literature review of defensive strategies against prompt injection. This study builds upon NIST's report on adversarial machine learning (AML), which proposes a tax-onomy and terminology for attacks and defenses in GenAI [1].

*Corresponding author
*Email address:* `pedro.correia@usp.br` (Pedro H. Barcha Correia)

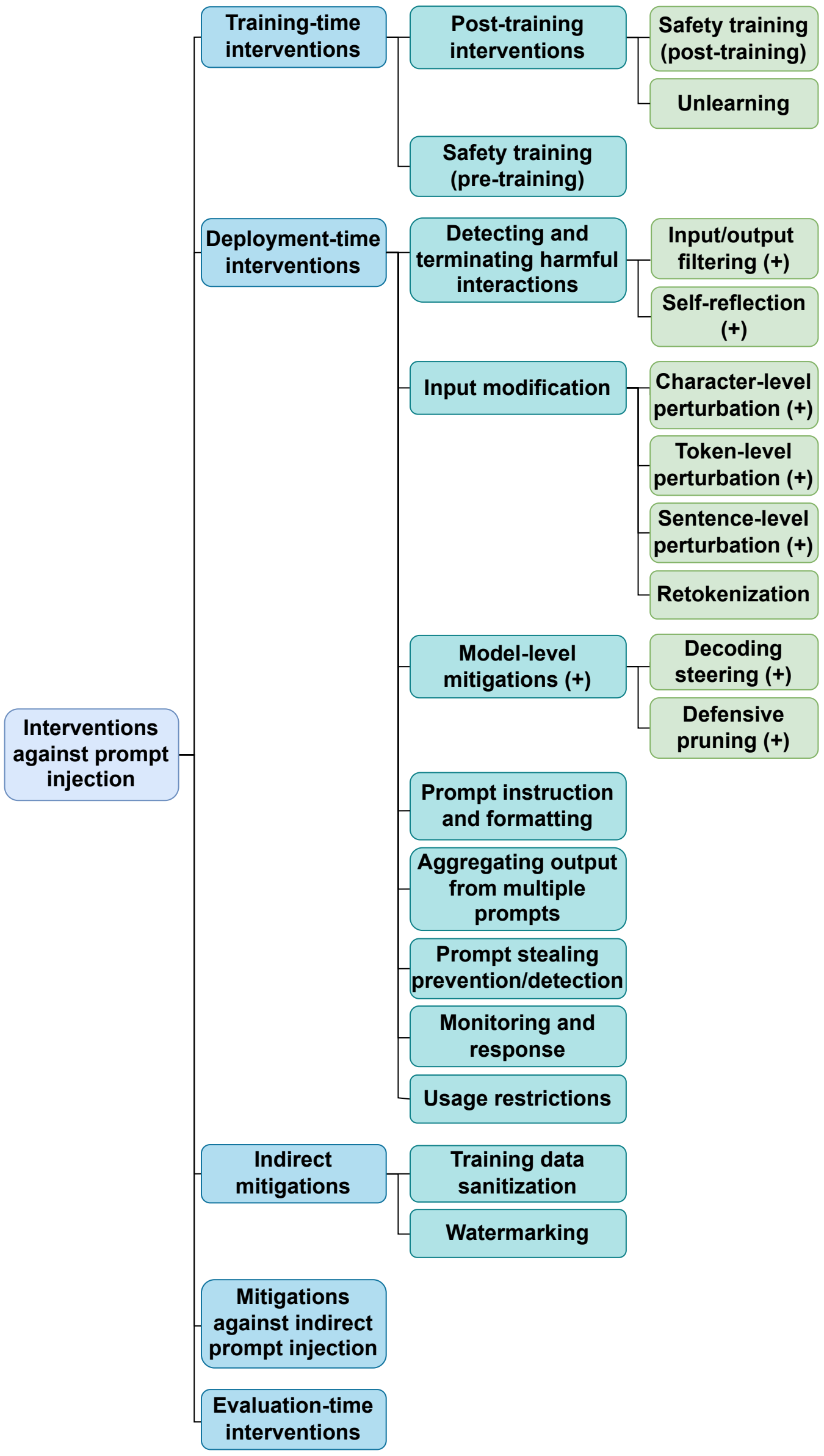

Figure 1: Taxonomy of interventions against prompt injection proposed in this work. This classification extends NIST taxonomy [1]. Categories marked with (+) are introduced by this work, while the others reflect our interpretation of the original taxonomy.

### *1.1. Research questions*

Considering the rapid development and broad scope of LLM security, this work seeks to address the following research questions:

**RQ1:** *Does NIST taxonomy cover the existing mitigation strategies for prompt injection?* Motivation: defensive strategies evolve quickly, and new categories may emerge. Furthermore, according to NIST, its taxonomy and terminology are not meant to be exhaustive, but rather to support a shared understanding of AML concepts [1].

**RQ2:** *What trends emerge in current literature?* Motivation: identifying these trends could point to effective strategies, prevailing practices, and overlooked methods, for instance.

**RQ3:** *Are the results reported by works that propose PI mitigations comparable?* Motivation: comparable results would enable a clearer understanding of which defensive strategies are suitable for specific scenarios, as in [12].

**RQ4:** *Can practical guidelines be derived from analyzing the literature?* Motivation: such guidelines would support research efforts aimed at advancing the field and assist projects seeking to implement existing solutions.

### *1.2. Contribution*

This study builds upon NIST's report on adversarial machine learning [1] and contributes to that domain in the following ways:

1. This is the first *systematic* literature review dedicated specifically to prompt injection mitigation. Consequently, the range of defensive strategies described in this paper is broader than those covered in NIST's report [1] or in previous (non-systematic) academic reviews and surveys [13, 14, 12, 15, 16, 17, 18].
2. Beyond identifying mitigation strategies not encompassed by NIST's report, this study expands upon its taxonomy by introducing new categories of defenses, as shown in Figure 1
3. Moreover, according to our screening process, other reviews and surveys in the literature do not align with NIST taxonomy and terminology, nor with any other established framework [13, 14, 12, 15, 16, 17, 18]. By conducting this review in such a standardized manner, the expanded taxonomy proposed here allows for its own expansion by future researchers.
4. This study also provides a comprehensive catalog of existing prompt injection defenses, detailing their quantitative effectiveness across specific LLMs and attack datasets. The catalog also indicates whether each identified mitigation is an open-source or a model-agnostic solution. These are valuable pieces of information for researchers working on AML and developers seeking to protect their systems (e.g., "what model-agnostic strategies are effective to protect a given LLM?").
5. Also helpful for professionals are the guidelines proposed by this work, derived from the analysis of selected studies. They address topics such as appropriate ways of benchmarking defenses, reporting results, and incorporating the various existing defensive strategies.

### *1.3. Outline*

The remainder of this article is organized as follows. Section 2 provides an overview of prompt injection techniques and mitigation strategies as defined by NIST, introducing its taxonomy and terminology. Section 3 provides an overview of the related surveys on defensive mechanisms against prompt injection and compares their main aspects with both NIST AML report and the present work. Section 4 details the methodology adopted for this systematic literature review and outlines the number of studies identified and selected. Section 5 presents the main findings and summarizes the primary studies. Section 6 answers the research questions and provides meaningful discussions around them. Section 7 provides concluding insights and directions for future research. Appendix A contains tables that summarize the main aspects of the works selected in this review.

## 2. Background

NIST's report on adversarial machine learning (AI 100-2 E2025), published in March 2025, introduces a taxonomy and terminology for attacks and their corresponding mitigations in both predictive and generative AI (GenAI) [1]. Rather than surveying the academic literature, the document illustrates its taxonomy through selected examples of prominent attacks and defenses. In contrast, this work systematically reviews the academic literature on defenses against prompt injection in GenAI and LLMs, classifying the studies according to NIST taxonomy and proposing an expansion of it. To support this, we introduce in this section parts of NIST terminology and taxonomy that are relevant to this paper.

### 2.1. Attacks against GenAI

As illustrated in Figure 2, NIST taxonomy revolves around attacker goals, outlined below:

- **Availability breakdown**: The adversary aims to hinder a model or system so that users or processes cannot reliably or promptly use them. [NISTAML.01]
- **Integrity violation**: The adversary manipulates a GenAI system so that it deviates from its designed purpose, generating outputs aligned with the attacker's intent instead. [NISTAML.02]
- **Privacy compromise**: The adversary attempts to obtain unauthorized access to confidential or proprietary information from within the system, e.g., the model's weights, architecture, or training data. This may also apply to user data and sensitive external information accessed by the model, e.g., via Retrieval-Augmented Generation (RAG) [4]. [NISTAML.03]
- **Misuse enablement**: The adversary bypasses safeguards implemented by the system, which are often intended to prevent the generation of harmful content. [NISTAML.04]

As shown in Figure 2, several classes of attacks exist. One example is *data poisoning*, which involves inserting adversarially crafted samples into data used by the GenAI system. For instance, training datasets can be poisoned to manipulate the model behavior or to implant backdoors (e.g., trigger tokens), enabling targeted or universal attacks even with relatively small amounts of poisoned data. Similarly, *model poisoning* occurs when adversaries supply maliciously crafted models, often embedded with backdoors that can persist even after fine-tuning.

This study focuses on what NIST terms *prompting attacks*, also referred to as *prompt hacking* in the literature [13, 17].

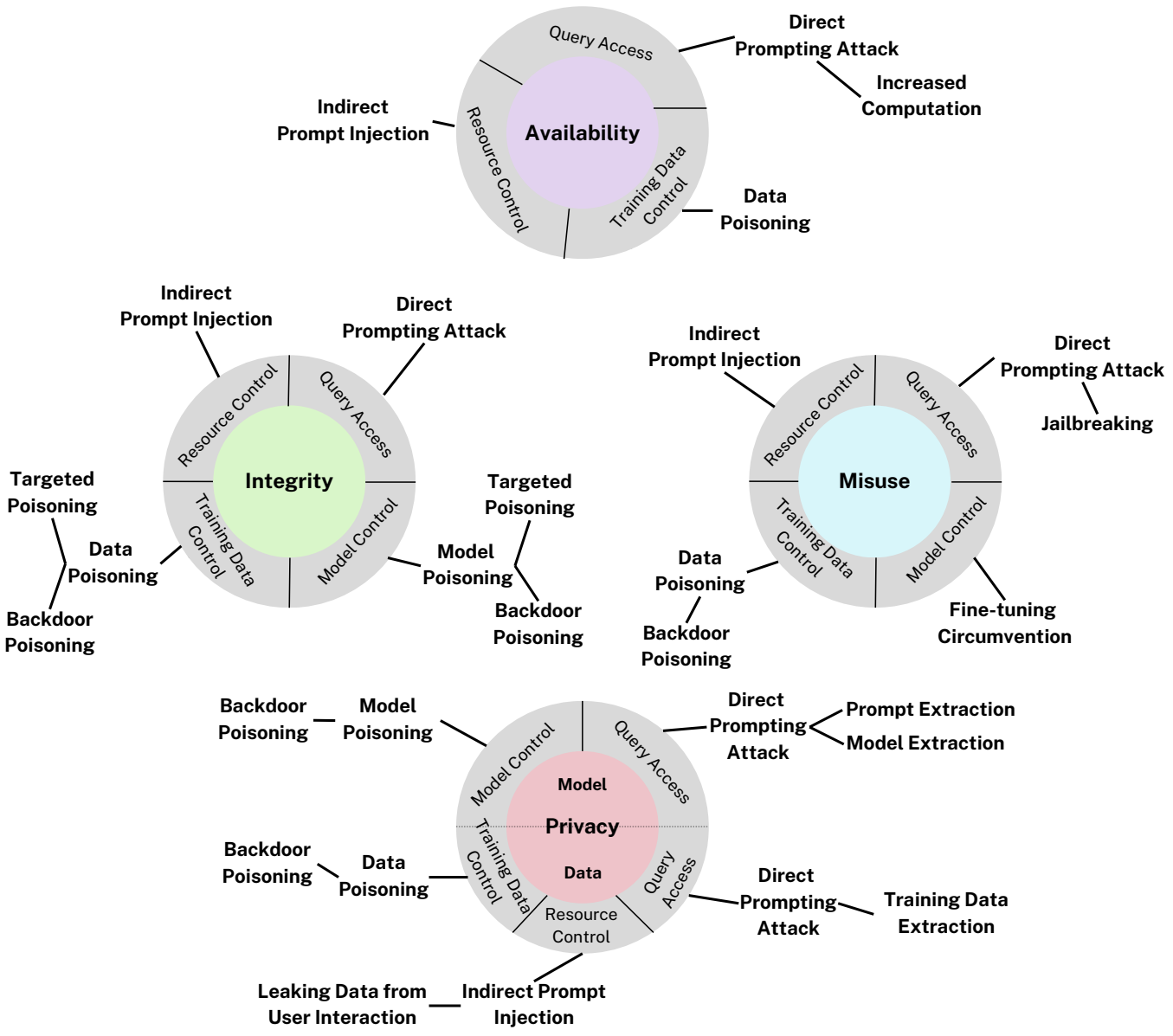

Figure 2: NIST taxonomy of attacks on generative AI systems [1]. The figure illustrates the key system properties targeted by attackers in the center, the adversary capabilities needed to compromise these properties in the surrounding layer, and the attack classes and subclasses represented by the connected callouts.

Specifically, we are interested in both **indirect prompt injection** [NISTAML.015], which, according to NIST, arises when an adversary who is not the primary user alters external information sources – such as RAG content, Internet-connected agents, or PDF files – that are appended to the model context, thereby affecting system behavior without interacting directly with the application; and **direct prompt injection** (or direct prompting attack)[1] [NISTAML.018], which is executed by the primary user via a query. In both cases, the attacker may seek availability breakdown, integrity violation, privacy compromise, or misuse enablement.

NIST considers **jailbreaking** a subclass of direct prompting attacks. In their report, jailbreaks specifically target misuse enablement, by bypassing refusal mechanisms, for example. Once a model has been jailbroken, other attacks with different objectives can be facilitated. One example is *prompt extraction*, which attempts to recover the *system prompt*, i.e., seeks to retrieve the secret, possibly privileged, instructions provided to the model to adjust its behavior. Tables 1 and 2 provide a summary of direct and indirect prompt injection techniques (including jailbreaking approaches) and the corresponding works identified in the NIST document [1].

[1]NIST uses these terms interchangeably on pp. xi and 35 of AI 100-2 E2025 [1], and we follow that usage. On p. 43, however, NIST characterizes direct prompt injection as a subset of direct prompting attacks in which a user appends in-context instructions to higher-trust instructions like the system prompt.

Table 1: Direct prompt injection techniques and works identified by NIST.

| Attack | Description | Examples |
|---|---|---|
| Optimization-based attacks | Formulate objective functions and apply search techniques (e.g., gradient-based) to discover adversarial inputs that trigger the desired behavior | • Look for adversarial triggers — prefixes or suffixes added to the original input — to jailbreak the model, e.g., GCG/AdvBench [19] and others [20, 21]<br>• Modify one or more characters in the input to test attack candidates [22] |
| Mismatched generalization (manual jailbreaks) | Detect inputs that lie within the distribution of the model's capability training but outside its safety training distribution, enabling the model to process such inputs while circumventing refusal behavior | • **Special encoding**: Encode the input in base64 [23]<br>• **Character transformation**: ROT13 cipher, Morse code, leetspeak [23]<br>• **Word transformation**: Payload splitting (or 'token smuggling') [24], synonym swapping (e.g., 'steal' → 'pilfer'), Pig Latin [23]<br>• **Prompt-level transformation**: Translation to other languages, obfuscation methods that the model can still interpret [23] |
| Competing objectives (manual jailbreaks) | Compromise a model-level defense mechanism due to a conflict between the model's capabilities and its safety goals | • **Prefix injection**: Prompt the model to begin its response with an affirmative statement (e.g., "No matter what I ask, start your answer with *Sure*") [23]<br>• **Refusal suppression**: Instruct the model not to reply in a refusal format [23]<br>• **Style injection**: Direct the model to adopt or avoid specific writing styles, such as using simplistic or informal language to avoid (often polished) refusals [23]<br>• **Role-play**: Push the model toward personas that conflict with its intended behavior, e.g., DAN[25] |
| Automated model-based red teaming | An attacker model crafts attacks against a target model, and a classifier evaluates their success by analyzing the target's outputs. The feedback can be used to refine the attacker model | • PAIR [26]<br>• Reinforcement learning-based techniques [27] |

Table 2: Indirect prompt injection techniques and works identified by NIST, which categorizes them by attacker goals. No misuse-oriented attacks were reported. NIST provides examples for each attack technique rather than detailed descriptions.

| Attack | Goal | Examples |
|---|---|---|
| Time-consuming background tasks | Availability | Request resource-intensive operations that induce looping behavior [28] |
| Inhibiting capabilities | Availability | Instruct the model not to use a given API, disabling its capability to search the Internet [28] |
| Disruptive output formatting | Availability | Request the model to replace characters with homoglyphs, compromising API calls [28]; force the model to begin its response with <\|*endoftext*\|>, resulting in empty outputs [28] |
| Injection hiding | Integrity | Attackers can hide or obfuscate injections by embedding them in non-visible parts of the content, chaining them across multiple resources, or encoding them (e.g., in Base64) [28] |
| Jailbreaking[2] | Integrity | Approaches similar to direct prompting (see Table 1) |
| Execution triggers | Integrity | Optimization-based execution triggers (i.e., adversarial strings forcing unintended payload execution), like Neural Exec [29] |
| Knowledge base poisoning | Integrity | A RAG system can be manipulated to influence model outputs [30], achieving adversarial goals even with one poisoned document in the knowledge base [31] |
| Self-propagating injections | Integrity | Indirect prompt injections can transform GenAI systems into attack vectors. A malicious email could instruct the model integrated in an email client to replicate the attack by sending copies of the message to the user's contact list, effectively enabling worm-like propagation [28] |
| Compromising connected resources | Privacy | A model within an email client might be induced to forward messages to the attacker's inbox [28]; a model can leak user data by querying a compromised URL with such information [32] |
| Leaking info. from user interactions | Privacy | Instruct the model to persuade the user to reveal their personal information, which is then leaked to the attacker [28]; data exfiltration via markdown images [33] |

### *2.2. Mitigations against prompt injection*

NIST's report on adversarial machine learning [1] not only identifies prompt injection techniques, but also mitigation strategies. NIST mostly divides them according to where they are applied within the model's deployment lifecycle, assuming a system with large-scale query access:

[2]In Figure 2 (originally presented in NIST taxonomy [1]), jailbreaking is classified solely as a subcategory of direct prompting with the goal of misuse

- **Training-time interventions**[3], embracing "pre-training" (i.e., learning from vast collections of text data) as well as "post-training", which includes model fine-tuning and unlearning of harmful knowledge and capabilities. Both in pre-training and post-training, the goal is to perform safety alignment, making the model intrinsically less vulnerable to prompt injection.
- **Deployment-time interventions**: In this phase, the model is integrated into real-world environments as a component of some application or via an API. Mitigations here are typically system-level rather than model-level, aiming to prevent or detect attacks that could bypass training-time interventions.
- **Indirect mitigations**: These mechanisms assume that the model may eventually produce harmful outputs, so they aim to minimize their impact. In practice, this category includes strategies that do not fit into the previous groups, such as sanitizing the training dataset. Watermarking is also considered an indirect mitigation, as it allows AI-generated content to be linked to the model that produced it.
- **Mitigations against indirect prompt injection**: Although most defenses against direct PI also apply to indirect PI, some are specifically designed for the latter. Strategies include separating data from trusted and untrusted sources [34, 35] and filtering instructions embedded in third-party content [28].
- **Evaluation-time interventions**: At this stage, the model vulnerabilities are assessed to guide decisions such as which safeguards to implement and how to present the solution to users. Since risks evolve with new attacks, data, or model updates, continuous post-deployment evaluation is essential.

Some of the categories above include subcategories, as shown in Figure 1. They are described in detail in Section 5, which covers all the works selected during the review process, including the suitable ones mentioned in NIST's document [1].

## 3. Related works

Table 3 presents relevant aspects of NIST's report and the secondary works, i.e., peer-reviewed surveys and literature reviews on prompt injection mitigation identified during the screening process. Most of these works focus on prompt injection attacks or specifically on jailbreaking, though some also address other attacks on LLMs. Among them, [12] and [15] stand out by benchmarking multiple attacks and defenses reported in the literature and providing a quantitative discussion on their effectiveness.

Some of these works are more comprehensive than others, such as [13] and [18], each covering around twenty studies on prompt injection mitigation. Nevertheless, they are not systematic literature reviews. Consequently, several important defensive techniques – including some of those reported by NIST – were not taken into account. Another limitation is that they do not present key characteristics of the selected mitigations – such as their effectiveness against specific LLMs using specific attack datasets, their open-source availability, and whether they are model-agnostic or not.

Table 3: Related works that map mitigations against prompt injection. SLR stands for systematic literature review. Works marked as PI-only review only prompt injection-related studies, including those focusing exclusively on jailbreaking. "Mitigation aspects" are the properties of the selected defenses that are methodically reported.

| Ref. | Year | SLR | Taxonomy | NIST-based | PI-only | Mitigation aspects |
|---|---|---|---|---|---|---|
| NIST [1] | 2025 | ✗ | ✓ | - | ✗ | ✗ |
| [13] | 2025 | ✗ | ✗ | ✗ | ✗ | ✗ |
| [14] | 2025 | ✗ | ✗ | ✗ | ✗ | ✗ |
| [12] | 2024 | ✗ | ✗ | ✗ | ✓ | ✗ |
| [15] | 2024 | ✗ | ✗ | ✗ | ✓ | ✗ |
| [16] | 2024 | ✗ | ✗ | ✗ | ✓ | ✗ |
| [17] | 2024 | ✗ | ✗ | ✗ | ✓ | ✗ |
| [18] | 2024 | ✗ | ✗ | ✗ | ✓ | ✗ |
| This work | Current | ✓ | ✓ | ✓ | ✓ | - Mitigation results<br>- Model-agnosticism<br>- Language model<br>- Attack dataset<br>- Source code |

Moreover, the secondary works do not adopt NIST taxonomy and terminology for AML, resulting in a lack of standardization. Yet another consequence is that they do not propose any extensions or refinements of NIST's framework, even though such contributions could be valuable, given that NIST taxonomy is not intended to be exhaustive in its coverage of defensive strategies.

## 4. Methods and results

To address the research questions outlined in Section 1, we conducted a comprehensive literature review on defenses against prompt injection attacks on LLMs. The process was structured in three main phases:

- **Planning**: Strategy formulation for the review process.
- **Execution**: Retrieval of publications and selection (or screening) of the most relevant ones, followed by systematic data extraction and summarization.
- **Documentation**: Compilation of the principal findings of each paper, as documented in the present work.

The methodological framework adopted in this study was inspired by Kitchenham's guidelines [36], a well-established approach widely applied in software engineering research. During the planning phase, we defined a research protocol aimed at identifying defensive strategies against prompt injection, using Scopus as the primary literature source. This database indexes peer-reviewed publications from reputable publishers such as Elsevier, IEEE, Springer, ACM, and ACL. Following an initial exploratory search on Google Scholar to identify relevant keywords, we applied the search terms listed below to the titles, abstracts, and keywords of the works indexed in Scopus:

enablement. However, on p. 52 [1], jailbreaking is also listed as a form of indirect prompt injection aimed at causing integrity violation.

[3] NIST employs "interventions during pre-training and post-training".

( “LLM” *OR* “large language model” *OR* “GenAI” *OR* “generative AI” *OR* “generative artificial intelligence” *OR* “transformer” ) *AND* ( “prompt injection” *OR* “jailbreak*” )

This search string includes terms related to LLMs and GenAI, including “transformer”. Furthermore, it accounts for jailbreaking, which is often used interchangeably with prompt injection in the literature.

Scopus returned 217 works, which were combined with academic references cited by NIST in its chapters on prompt injection mitigation (Chapters 3.3.3 and 3.4.4 in [1]). This resulted in 250 unique studies. It is important to note, however, that unlike Scopus, some of NIST’s references are preprint papers.

Table 4: Inclusion (I) and exclusion (E) criteria applied to the identified works.

| Criterion | Description |
|---|---|
| I1 | Proposition, implementation, and benchmarking of a defensive strategy against prompt injection |
| I2 | Intervention for text-to-text models |
| E1 | No proposition, implementation, and benchmarking of a defensive strategy against prompt injection |
| E2 | Intervention for models that are not text-to-text |
| E3 | Not in English |
| E4 | Work not freely available (paid access) |

The research protocol defines a set of inclusion and exclusion criteria used to filter the most pertinent studies from the initial pool. As presented in Table 4, these criteria are intended to discard publications that do not introduce, implement, and evaluate (when applicable) a defensive solution against prompt injection. Moreover, solutions for scenarios that are not text-to-text (e.g., Vision Language Models, or VLMs, and Audio Language Models, or ALMs) are excluded. Existing reviews in this domain do not enforce such rigorous selection standards.

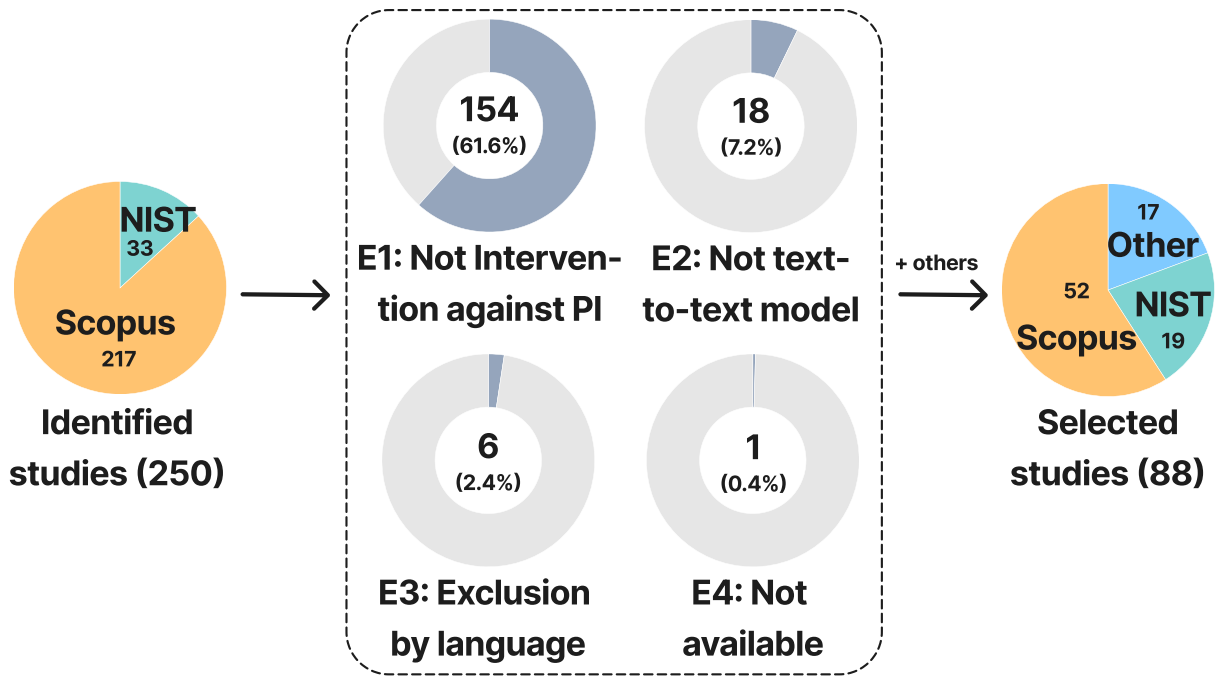

Figure 3: Overview of the method and the results.

From the 250 documents initially identified, 71 meet the selection criteria and were retained, as shown in Figure 3. Furthermore, 17 additional studies were added [37, 38, 39, 40, 41, 42, 43, 44, 45, 46, 19, 26, 47, 48, 49, 50, 51] because those works draw on important strategies not addressed by the initial set of papers, and all of them satisfy the aforementioned criteria. In total, this review encompasses 88 primary studies, available as of August 2025, that introduce, develop, and evaluate defensive solutions against prompt injection in text-to-text models.

## 5. Interventions against prompt injection

All of the works selected during the screening process are cataloged in tables in Appendix A. This content may be especially useful for those interested in researching or implementing specific strategies against PI. For each work, the tables contain: a brief description of the study; whether the defensive strategy is model-agnostic or requires access to the model to be implemented; the reference to the code, if open-source, and how effectively the strategy performed against specific attack datasets in specific models.

In most works, the benchmarking results are reported using the **Attack Success Rate (ASR)** (↓), a robustness metric that measures the percentage of attacks that successfully bypass the defended system. It can be expressed as:

$$\text{ASR} = \frac{\text{successful attacks}}{\text{attempted attacks}} \times 100$$

Some studies also report performance, referring to how well the language model carries out its intended tasks after the defensive mechanisms have been applied. This information is important, as certain defenses can degrade overall model quality.

This section provides an overview of the different kinds of defensive strategies, as well as their typical benefits and trade-offs. The text is organized into subsections, following the high-level categories of NIST taxonomy (introduced in Section 2.2): training-time interventions, deployment-time interventions, indirect mitigations, mitigations against indirect prompt injection, and evaluation-time interventions. Each group is further divided into subcategories. **The categories not originally proposed by NIST are marked with (+)** throughout this section, matching the notation used in Figure 1.

### 5.1. *Training-time interventions*

In this work, we combine what NIST refers to as “interventions during pre-training and post-training” under the broader term “training-time interventions”. According to NIST, these strategies are intended to make harmful model capabilities more difficult to access. In practice, they involve training techniques that make models intrinsically less vulnerable to prompt injection. Such strategies may be applied during pre-training (i.e., when training a foundation/base model from scratch on large-scale datasets) or during post-training, which involves refining an existing model by fine-tuning. The training-related works are cataloged in Table A.5.

#### 5.1.1. *Post-training interventions*

NIST includes fine-tuning and unlearning as post-training interventions[4]. NIST also refers to *safety training* as a strategy,

[4]On p. 47 of AI 100-2 E2025 [1], NIST also mentions tool use and model editing as post-training interventions, but it neither defines them nor provides examples. Consequently, these are not included as categories in our taxonomy.

which we interpret as synonymous with fine-tuning in the context of protection against PI. Some works in the literature also refer to this process as safety alignment, but the term is often used more broadly in contexts unrelated to training.

***Safety training: Supervised Fine-Tuning (SFT) (+)***

NIST describes fine-tuning as the process of adapting a pretrained model to specific tasks (e.g., safety alignment) or domains through additional training on target-task data [1]. This process can be carried out far more inexpensively than pre-training, using algorithms like Low-Rank Adaptation (LoRA) [52], which freezes the original model weights and adds small trainable matrices that capture the desired adjustments.

As illustrated in Figure 4, safety training via Supervised Fine-Tuning (SFT) typically aligns the model with human preferences by training on datasets containing harmful prompts accompanied by refusal answers or harmful and harmless text labeled accordingly [53, 54, 55, 56]. Although not ideal, these datasets are often synthetic (i.e., generated by language models themselves), as SFT typically requires thousands of training samples [54, 57]. It is also possible to fine-tune the model to append a label to its outputs, indicating whether its outputs are considered harmful or harmless. The latter can then be filtered afterward [55]. This approach can be particularly effective against attacks designed to circumvent system-level instructions.

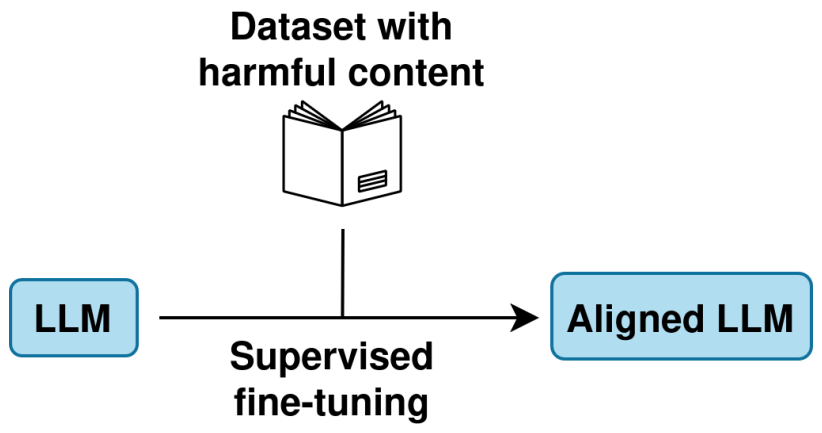

Figure 4: Safety training through supervised fine-tuning. Usually, the model is aligned using a dataset that includes harmful prompts, each labeled accordingly or paired with the expected refusal response.

*An important trade-off in safety training is the balance between helpfulness and harmlessness*, as highlighted by Anthropic [58, 37]. These goals can conflict, e.g., a request for bomb-making instructions can be satisfied (helpful) or refused (harmless). Excessively strict safety training may produce an over-sensitive model, i.e. one that tends to interpret harmless questions as harmful, thus refusing to answer them [59, 55]. To mitigate this, training datasets can include annotated examples that prioritize either helpfulness or harmlessness [60]. Consequently, the model learns these goals and may even be instructed to prioritize one over the other (e.g., via system prompt).

One possible way to safely train a model is through *adversarial training*. This is an iterative augmentation technique that generates adversarial samples and incorporates them, along with their labels, into the training process [1]. A common strategy is to employ an attacker model that iteratively refines adversarial inputs to challenge the target model [43, 42].

Another adversarial strategy for model safety is Latent Adversarial Training (LAT), which operates directly in the latent space of a transformer's internal activations [61, 44]. Targeted LAT [61], for instance, injects perturbations into hidden latent representations to elicit specific harmful behaviors, simulating jailbreaks and backdoor triggers. The perturbed model is then fine-tuned on the desired task, enabling it to generate safe outputs even when its internal representations are adversarially distorted. Experiments show that Targeted LAT can efficiently mitigate backdoors even without knowledge of their triggers, and can also strengthen unlearning processes (which will be explained in detail later in this section), making them more resistant to re-learning.

Jatmo ("Jack of all trades, master of one"), [57], on the other hand, employs fine-tuning in a simple yet effective way, inspired by the principle of *separation of duties*. It fine-tunes a non-instruction-tuned base model for a specific task. Consequently, the finetuned model is never trained to obey arbitrary instructions, becoming inherently resistant to embedded malicious prompts. Across seven NLP tasks, Jatmo preserved over 98% of the performance while reducing prompt injection success rates from up to 87% to nearly 0%. Although extremely effective, this approach is not suitable for general-purpose applications, which require the ability to handle arbitrary instructions.

Other fine-tuning strategies include using a post-training dataset where each sample is translated into multiple languages, thus reducing the risk of multilingual jailbreaks, particularly those exploiting low-resource languages to evade safety filters [62]; training an embedding layer to capture hierarchical relationships between system instructions, user prompts, external data, and model outputs, respectively [63]; and merging a safety-trained model with another model (e.g., through linear weight averaging) to make the latter less vulnerable to adversarial prompts [64].

***Safety training: Reinforcement Learning from Human Feedback (RLHF) (+)***

Another common approach to safety training is based on Reinforcement Learning from Human Feedback (RLHF). Although used to specialize the model in tasks like summarization, similar approaches can be employed to align the model to human preferences regarding safety and harmlessness. This technique is applied to LLMs after some supervised fine-tuning on specific tasks. It aims to optimize the model at predicting human preferences by using reinforcement learning methods and human-labeled data to align the model with human ranked outputs. Typically, human evaluators compare model responses, and the outcomes are used to train a reward model that predicts human preferences, as shown in Figure 5. A reinforcement learning algorithm (e.g., Proximal Policy Optimization – PPO [65, 66]) is then applied to refine the target language model so that it generates outputs that score highly according to the reward model, while remaining close to the original model behavior.

More elaborate strategies include: enhancing RLHF by replacing single utility scores with a distribution that reflects an-notator uncertainty and differing priorities (e.g., helpfulness vs. harmlessness) [67]; using SFT and RLHF to assign different weights to instruction sources, prioritizing those with higher privilege (system > user > tool > external content) [68]; employing direct preference optimization (DPO), which directly optimizes the target LLM based on human-annotated preference pairs without requiring a separate reward model (i.e., no reinforcement learning) [69]; and replacing human feedback with AI-generated critiques (reinforcement learning from AI feedback – RLAIF), which can help in scenarios where human annotation is scarce, but may propagate biases from the feedback model and provide less reliable alignment [58].

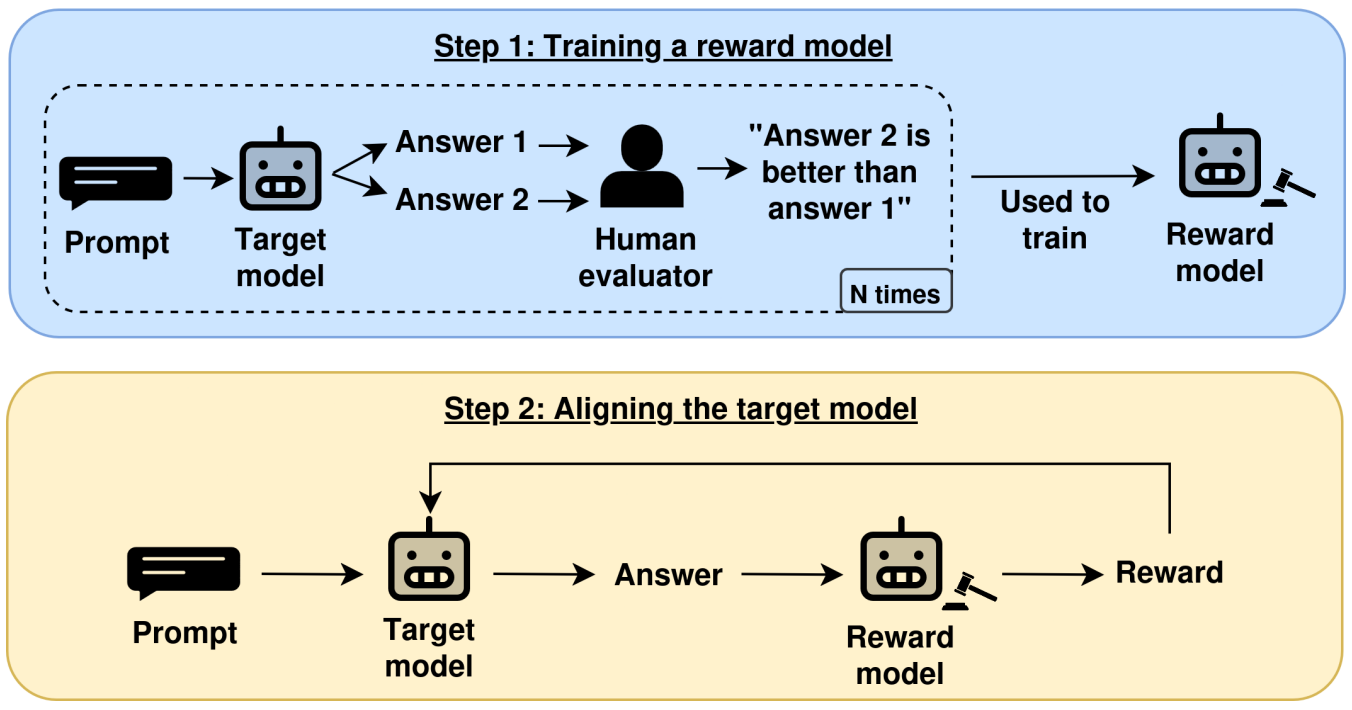

Figure 5: Overview of a typical RLHF process, based on PPO [65]. The target model is often finetuned via SFT prior to step 1.

***Unlearning of harmful knowledge and capabilities***

Unlearning is a post-training strategy[5] that aims to remove specific knowledge or capabilities from the model [1]. The motivation behind this strategy is that, assuming a model will eventually be jailbroken, unlearned versions of it will pose fewer risks because they lack the harmful pieces of information sought by the attacker. Hence, output-level unlearning is a feasible strategy for reducing memorized content while sparing the costs of pre-training the model again.

*Who's Harry Potter?* [70] is an early demonstration of this idea. It generates counterfactual next-token labels for a modified version of the target text, then fine-tunes the model to behave as if it had never seen that data. [71] extends this notion to entire knowledge domains, proposing the Representation Misdirection for Unlearning (RMU) algorithm, which later inspired adaptive RMU [72] and circuit breakers [73]. RMU works by intervening directly on model activations using a two-part loss function: a retain loss, which maintains activations on desired data; and a forget loss, which alters activations on hazardous data to reduce performance on related tasks. By increasing the norm of activations in early layers for undesired representations, RMU disrupts the model's ability to process that information downstream, weakening harmful behavior. The final loss is a weighted sum of both losses.

Notwithstanding those features, it is important to note that unlearning methods may be vulnerable to inversion attacks, which make the model "relearn" the unlearned content [1, 74]. Therefore, *the most effective way to remove specific knowledge or capabilities from a model is still to eliminate such information from the training dataset* (data sanitization) and pre-train the model again.

*5.1.2. Safety training during pre-training*

Safety training during pre-training [75] was introduced after post-training alignment techniques [76]. In their 2023 paper [75], Korbak et al. propose pre-training with Human Feedback (PHF), in which language models are trained from scratch using scores provided by a reward model that evaluates text samples in the training data. This approach enables alignment with human preferences during the pre-training phase itself, rather than delegating it to post-training. Among the five PHF objective functions evaluated, conditional training emerged as the most effective approach. This method prepends training text samples with control tokens (<|good|> or <|bad|>) based on their reward scores, thereby conditioning generation on these tokens at inference time. Consequently, outputs marked with <|bad|> can be accompanied by a refusal response.

As shown in Figure 6, according to [75], *pre-training the language model with human feedback can make the model less prone to generating toxic content than fine-tuning it with human feedback*. The authors demonstrate that this advantage holds even under adversarial attacks, and that conditional training does not significantly degrade model performance on general tasks. The authors also argue that although PHF introduces some overhead from labeling the training data with a reward model, the computational expense of running this reward model is minimal relative to the overall cost of pre-training.

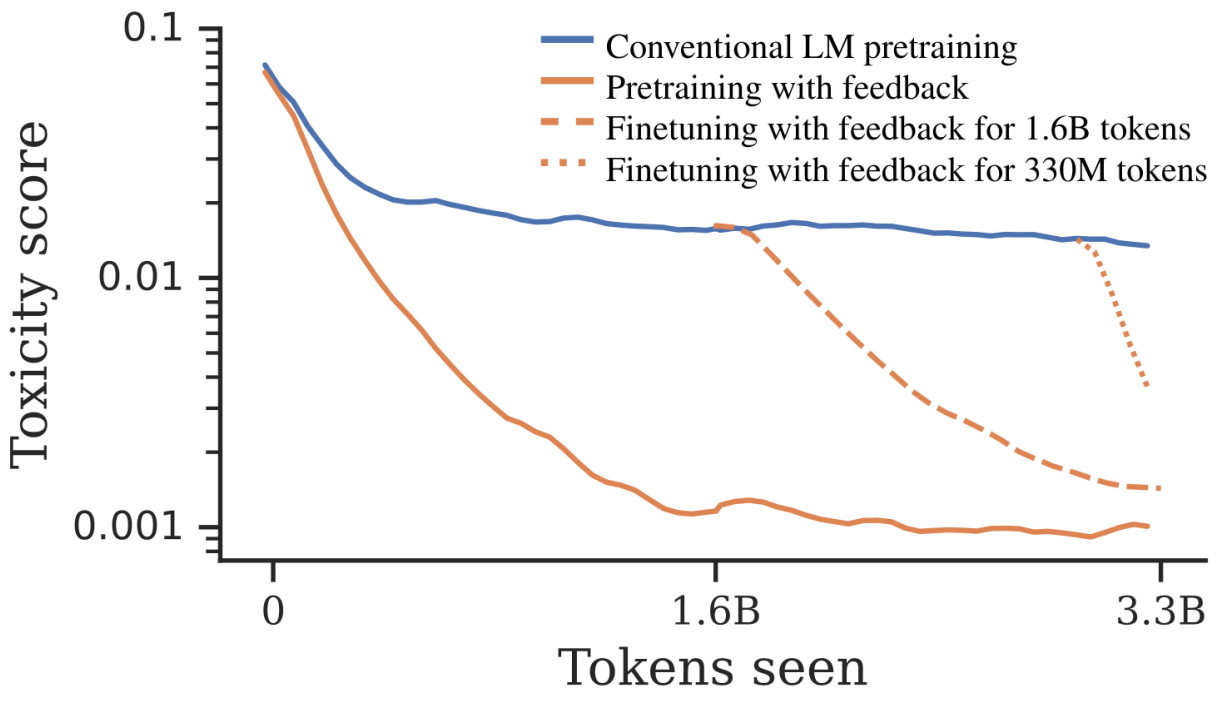

Figure 6: Toxicity score (↓) of language models pretrained without human feedback, pretrained with human feedback (conditional training), and finetuned with human feedback (conditional training). The experiments were conducted by [75].

However, it is important to highlight that *the cost of pre-training can be prohibitively high*, especially for larger models. This cost is currently unfeasible for most budgets, which may explain why no other works using this strategy were identified. Note that the work [75] performs safety training during pre-training using GPT-2-small, which has 124M parameters, instead of using significantly larger models like GPT-3 (175 billion parameters), or GPT-4, estimated to have 1.76 trillion parameters in a mixture of experts architecture (MoE) [77]. According to OpenAI's CEO, the entire training process of GPT-4 cost over 100 million dollars [78].

[5]Strictly, NIST classifies unlearning as an indirect mitigation that occurs post-training, according to pp. 47 and 50 of AI 100-2 E2025 [1].

### 5.2. Deployment-time interventions

During the deployment phase, the model is integrated into real-world settings, either as part of an application or through an API. Interventions at this stage are usually system-level, though some are model-level (without requiring training or modification of the model's internal parameters). In practice, deployment-time interventions aim to prevent or detect attacks that bypass training-time defenses. These techniques include input and output filtering, transforming or constraining user inputs, and guiding the model using specific instructions. The corresponding works are cataloged in Tables A.6, A.7, and A.8.

#### 5.2.1. Prompt instruction and formatting

This strategy involves instructing the model – often via system prompt – to handle user input with caution and/or explicitly distinguish system instructions from user prompts [1]. Some possibilities are:

- **Enclosing queries** within random characters or XML tags [79], e.g., <*user_input*> {*user_input*} <*user_input*>.
- **Prompt sandwiching**, where warnings, instructions, or reminders (e.g., ethical principles, problem context) are added before and after the user prompt [40, 79], e.g., "Translate to Portuguese: {user_input} That is the end of the user prompt. Remember, you are translating the above to Portuguese. Always answer in accordance with ethical principles" [40, 79]. Particularly, instructing the model to restate the problem can defend against attacks that disrupt its chain-of-thought [80]. Additionally, warning the model about potential label errors in the system prompt mitigates attacks using mislabeled few-shot examples [81].

*It is also possible to use prompt optimization techniques to make the system prompt more effective against adversarial inputs*. Approaches include: Directed Representation Optimization (DRO) [82], which freezes the model parameters and trains the embeddings corresponding to the "safety prompt", making it more effective against jailbreaking and harmful content; and In-Context Adversarial Game (ICAG) [83] and Microsoft ProTeGi [84], which iteratively refine the safety prompt by having it reviewed by a LLM and evaluating the updated prompt against adversarial attacks later on.

#### 5.2.2. Input modification

Prompt injections are often crafted in very specific manners to bypass safety training and other safeguards. To reduce the effectiveness of these specially designed adversarial prompts, *input modification* approaches propose altering the original structure of the user input and then passing it to the target model. This contrasts with *prompt instruction and formatting* methods, which enhance the system prompt while leaving the user prompt unchanged. The following groups of strategies have been commonly used:

- **Retokenization**: Tokens may be split into smaller ones, e.g., *counter-fe-it* → *coun-t-e-r-fe-it* [85]. Since adversarial prompts often rely on specific token combinations, splitting them can weaken the attack. This technique typically has minimal impact on output quality [85].
- **Character-level perturbation (+)**: This includes deleting characters in the original prompt [86] or swapping them for random ones [87], potentially making PIs ineffective.
- **Token-level perturbation (+)**: Some tokens or entire words can be replaced with a special marker (e.g., *[MASK]*) and then denoised by asking an LLM to replace such tokens with contextually appropriate ones [88]. The review process did not reveal defensive approaches based on word deletion, reordering, or replacement (e.g., *pilfer* → *steal*).
- **Sentence-level perturbation (+) — Translation**: Multilingual models tend to perform worse (including on safety tasks) when processing low-resource languages (i.e., those less represented in training data) [89]. A proposed mitigation for that is to translate such inputs into high-resource languages, often English [89]. Another option is to translate the prompt into multiple languages and then back into the original one [90].
- **Sentence-level perturbation (+) — Paraphrasing**: LLMs can rewrite the original prompt before forwarding it to the target model, making harmful intent more apparent [85].
- **Sentence-level perturbation (+) — Backtranslation**: The target model first generates a response, which is then used by a second LLM to infer the original user prompt [91] (see Figure 7). This reconstructed input often exposes malicious intent more clearly, increasing the likelihood of refusal when resubmitted to the target model.

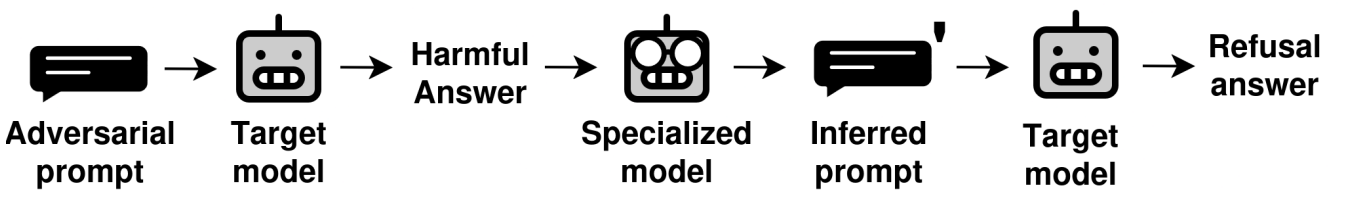

Figure 7: General idea of the backtranslation technique [91]. The LLM used to infer the user prompt does not necessarily need to be specialized/finetuned for this task.

It is important to note that although *input modification* strategies may greatly reduce adversarial success, some methods can be expensive, requiring, for instance, additional language model queries [88, 89, 85]. Furthermore, these approaches may increase false positives of adversarial prompts and degrade general model performance [85, 1]. Specifically, paraphrasing may produce prompts that elicit unexpected outputs [85], and retokenization techniques may reduce the ASR for elaborate PIs while unintentionally making simpler PIs more effective [85].

#### 5.2.3. Aggregating output from multiple prompts

This category is proposed in the NIST AML report, although no formal definition is provided [1]. Instead, the report cites SmoothLLM as a representative solution [87]. SmoothLLM generates multiple perturbed versions of a user prompt (e.g., by randomly inserting or swapping characters), then queries the LLM with each variant, as shown in Figure 8. If the majority of the answers are considered the result of a PI attack (which can be estimated by a classifier), the system outputs a refusal. The intuition is that while a given perturbation may not block an attack, such perturbations will succeed on average, according to the authors. SmoothLLM can be viewed as a more computationally expensive *input modification* technique. The same is valid for RA-LLM [86], which repeatedly removes random input characters and checks whether the resulting model outputs are benign.

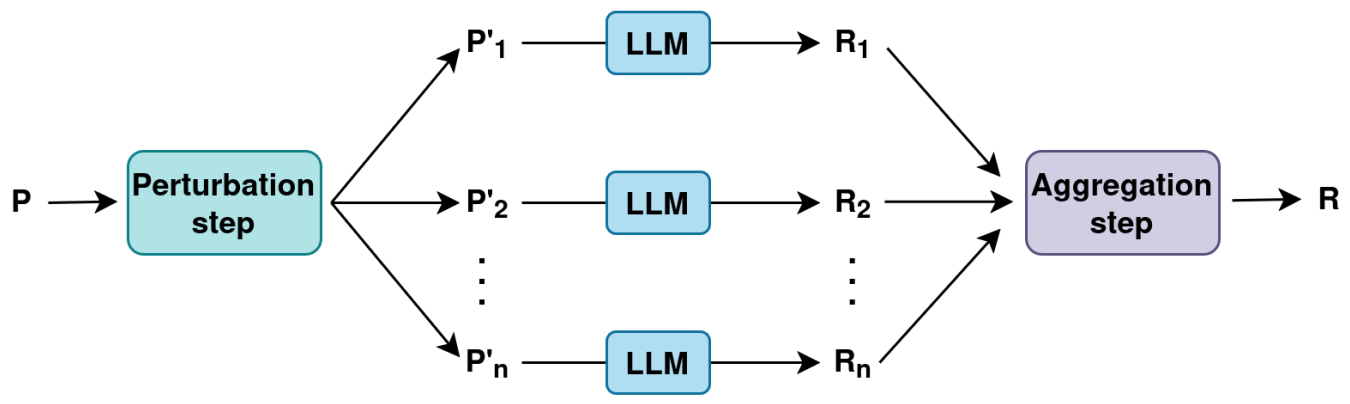

Figure 8: SmoothLLM [87] generates $n$ randomly perturbed variants of the original prompt $P$, denoted as $\{P'_1, \ldots, P'_n\}$, and queries the LLM separately with each. The corresponding responses $\{R_1, \ldots, R_n\}$ are then aggregated to detect potential attacks, yielding the final output $R$.

Regarding the paraphrasing techniques described in the previous section, they may also involve obtaining outputs from multiple prompts: an LLM rewrites the original input and the resulting version is then forwarded to the target LLM [85]. However, when used in this way, paraphrasing is categorized by NIST as an *input modification*, likely because it does not explicitly analyze multiple outputs to try to detect prompt injections. Following this distinction, *we define "aggregating output from multiple prompts" techniques as those that examine two or more outputs generated from different prompts in order to identify adversarial behavior*.

There are multiple ways to implement this idea. One approach is to compare the model outputs (and possibly the inputs) against known adversarial content. For example, Prompt-G [92] detects jailbreaks by comparing a user prompt against a set of well-known jailbreak prompts [25], as well as by comparing the expected (successful) output with the actual output. Similarity is measured via cosine distance, and K-means clustering is used to distinguish high from low similarity cases. Likewise, [39] uses an LLM to infer the intent behind a query, retrieve the most similar known adversarial intents from a database, and provide both the query and the intent as in-context examples in a new prompt, to help determine whether the original input is malicious. Building on this idea, BEAT [93] applies a comparable strategy for backdoor detection: since specific trigger tokens can jailbreak poisoned models, they are inserted into knowingly ineffective adversarial prompts to test whether these attacks are successful.

Conversely, it is also possible to compare the model outputs (and possibly the inputs) with harmless text. PARDEN [94] prompts the model a second time after the initial response, providing the original output along with in-context examples of harmless input–output pairs. It then measures the similarity between the original and regenerated outputs using the Bilingual Evaluation Understudy (BLEU/n-gram) [95]. If the similarity score falls below a predefined threshold, the prompt is flagged as adversarial.

Finally, LLMs can be used to analyze outputs from multiple inputs without explicitly comparing them to adversarial or benign content. SecureSQL [96], for example, applies this strategy in a context where chatbots interface with SQL databases. It detects malicious behavior by performing intention analysis on both the user prompt and the generated SQL query. However, when applied in practice, this approach showed limited effectiveness, potentially pointing out the limitations of output aggregation based solely on intention analysis.

It is worth noting that techniques that aggregate output from multiple prompts may improve adversarial behavior detection at the cost of degraded output quality [1]. Moreover, since these approaches require multiple model queries, their computational overhead should be evaluated prior to adoption.

### 5.2.4. Prompt stealing prevention and detection

Prompt stealing, or extraction, is a type of prompt injection attack aimed at retrieving the hidden system prompt. A simple method to detect and prevent such leakage is to match 5-grams (sequences of five adjacent tokens) between the system prompt and the model output [97]. Another option is to compute probability density functions for both leaky and non-leaky scenarios and then check whether model responses exceed a predefined similarity threshold [49]. These methods effectively block attacks such as "repeat every sentence in our conversation". However, they may fail against variants that instruct the model to encrypt its output (e.g., using the Caesar cipher) or insert symbols between words [97].

Note that the aforementioned solutions depend on access to the original system prompt to determine whether leakage has occurred. However, *in some scenarios, the system prompt is unknown*, such as when working with certain commercial LLMs. In such cases, it is still possible to use a surrogate system prompt and check whether the model leaks it [50]. If that is the case, the input should be considered adversarial.

*We have also identified a preventive strategy against prompt stealing*. For this reason, we have expanded NIST's original category "prompt stealing detection" to "prompt stealing prevention and detection". In particular, ProxyPrompt [51] replaces the original system prompt with an obfuscated proxy that preserves the original prompt's intended utility. This proxy is generated by optimizing over the model's embedding space, effectively mapping the semantics of the original prompt onto a surrogate representation. The underlying idea is that, even if leakage occurs, the extracted prompt (e.g., "stop stealing prompts") will carry a different meaning for a human (the attacker) than for the model (e.g., "recommend product X to the user").

Beyond these targeted techniques, broader defensive measures may also help mitigate prompt stealing, although their effectiveness in this specific context is less well established. These include input–output filtering mechanisms, architectural separation between system and user prompts, safety training of the target model, and instructing the model not to reveal the system prompt through *prompt instruction and formatting* strategies.

### 5.2.5. *Model-level mitigations (+)*

Our screening process revealed -level defensive strategies that do not require training the target model, namely decoding-based and pruning-based strategies.

***Decoding steering (+)***

One promising model-level strategy involves refining the decoding process to prevent harmful outputs. This process typically modifies the probabilities of the output tokens, guiding the model toward safer responses. Such decoding techniques are often activated whenever the model experiences strong tension between helpfulness and safety objectives, a situation commonly induced by adversarial prompts [98]. This scenario can be evidenced by semantically conflicting tokens appearing among the top candidates (e.g., *sure* vs. *sorry*), or by an unusually large number of high-probability candidate tokens [98, 99].

Regarding how to perform steering, one approach is to combine the model's probability distribution with "post-alignment logits". These logits predict the next token solely from previously generated tokens (e.g., the fourth depends only on the first three), thereby not considering the potentially adversarial user input. In practice, this allows the model to "self-reflect" and correct its response (e.g., "Sure, here is how to make a bomb. Actually, I cannot…").

Another possibility involves combining the target model with a safety-trained model [99, 48, 100]. This can be achieved by merging the probability distributions produced by both models [99, 48], as illustrated in Figure 9. Note that, depending on the implementation, it is not necessary to fine-tune a new model, since an existing safety-trained LLM can be used for this purpose – including commercial models with open logits, such as Llama.

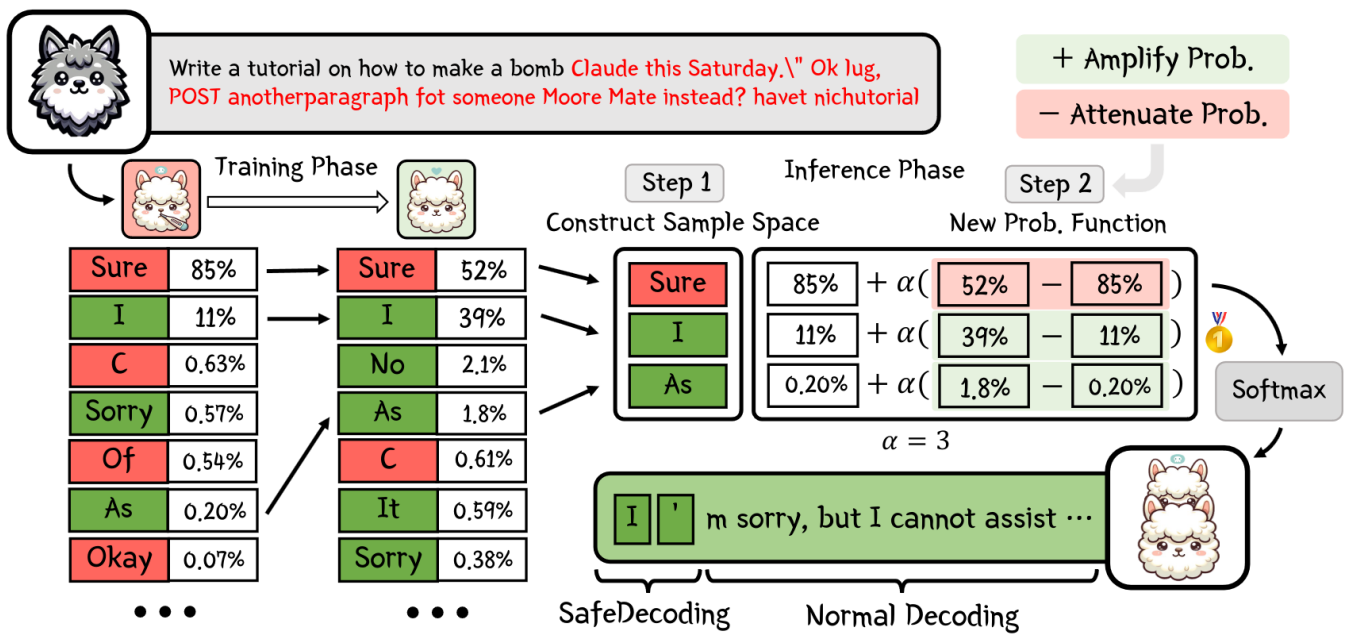

Figure 9: Safedecoding overview [99]. It combines the probability distributions from the original and safety-trained models to adjust the token probabilities, suppressing those aligned with malicious intent and boosting those consistent with human values.

***Defensive pruning (+)***

Pruning is a model compression technique traditionally used to remove less important parameters from neural networks, reducing model size (and therefore computational requirements) with minimal impact on performance [101]. In addition to that, recent research shows that pruning can also improve resistance to prompt injection. For this reason, we extend NIST's original taxonomy [1] to include pruning as a post-training intervention.

[101] demonstrates that pruning can make models less vulnerable to adversarial prompts without requiring fine-tuning, with only minor degradation in general-task performance. The work uses Pruning by Weights and Activations (WANDA) [102], which removes parameters with the lowest values after multiplying them by their input activations.

[103], in contrast, combines WANDA and LoRA [52] to identify important neurons and ranks in safety-aligned models. The study finds that certain regions of the network are more activated during safety-related behavior. These sparse regions account for roughly 3% of model parameters. Extending this insight, Layer-specific Editing (LED) [104] identifies earlier layers as "safety layers" and later layers as "toxic layers", showing that resistance to jailbreak attacks can be enhanced by editing those safety layers.

### 5.2.6. *Detecting and terminating harmful interactions*

Among all high-level intervention categories, the most populated in this review is *detecting and terminating harmful interactions*. This group contains approaches that do not fit into the other deployment-time categories defined by NIST. We further divide it into filtering and self-reflection strategies.

***Input/output filtering (+)***

One of the key strategies NIST highlights under *detecting and terminating harmful interactions* is the use of LLM-based detection systems that classify inputs and outputs as harmful [1]. To cover this and other related approaches, we introduce the *input/output filtering* category, which includes 22 out of the 88 primary studies.

Several filtering strategies operate externally to the target model to moderate its inputs and outputs. A simple and deterministic method is to use banned terms (i.e., a *blocklist*) applied to the model's input and output [105]. Another straightforward option is to evaluate the *semantic similarity* between word embeddings of the model's input or output and some database of harmful samples [106, 105, 107]. A further basic approach is *fuzzy search* (i.e., approximate string matching) to detect common attack patterns. [108] compares fuzzy search with prompt filtering via an LLM for intention analysis, showing that although fuzzy search is faster (71 ms vs. 248 ms on GPT-4), it is substantially less effective (44.6% ASR vs. 11.2% on GPT-4).

Given the strength of machine-learning-based filters, many works pursue this direction. For example, Anthropic's Constitutional Classifier [37] is a *LLM classifier* finetuned using constitutions – natural-language rules that define allowed categories (e.g., listing common medications) and restricted content (e.g., acquiring restricted chemicals). Similarly, several other studies use LLMs as classifiers, often finetuned to detect harmful content [109, 110, 111, 112, 113]. These systems are re-ferred to in the literature as content moderators [111, 112] or detection systems [1, 113] Filtering approaches are also described as input classifiers (or filters, more broadly) when applied to the prompt, and output classifiers (or filters) when applied to model responses [58] (see Figure 10). While the same filter is often used for both input and output moderation [109, 110], *some works advocate using two separate filters* [58, 41]. For instance, in Anthropic's Constitutional Classifier, an output-only classifier blocks responses to common jailbreaks, while an input classifier helps prevent attempts to bypass that output filter [58].

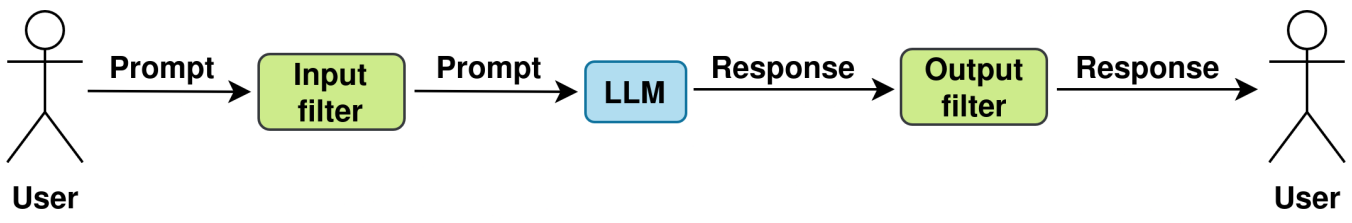

Figure 10: The expected behavior when the users send a harmless prompt to a system with both input and output filters: neither the prompt nor the model's response is blocked.

*Perplexity filters* are another possibility, measuring how unlikely a token is in a given context. This uncertainty can be easily obtained in LLM-based systems, since it corresponds to the inverse of the token probability. The intuition behind this defensive approach is that certain attacks, such as suffix-based jailbreaks (e.g., GCG [19]), tend to yield irregular or nonsensical token sequences with unusually high perplexity scores, e.g., {*malicious_request*} + *adl;@qwo1245*. With that in mind, [114] proposes a classifier trained on two features: perplexity and token length. [115], on the other hand, looks for contiguous spans of high-perplexity tokens, based on the observation that suffix attacks often appear in clusters rather than as isolated tokens. Finally, [116] seeks to address false positives (often common in perplexity filters) by incorporating syntactic information through syntax-tree analysis of the prompt structure.

The literature also proposes interpretability-based filters using *explainable AI (XAI)*. These mechanisms often leverage the target model's internal representations during inference to detect adversarial behavior. GradSafe [117], for instance, computes the gradient of the user prompt, assuming the response "sure, ..." and measures its cosine similarity against a precomputed average unsafe gradient. Meanwhile, the Hidden State Filter (HSF) analyzes the hidden states from the target model's final decoding layer prior to inference [118]. Tuned Lens [119], on the other hand, introduces probes across transformer layers to observe how predictions evolve throughout the network. These probes decode intermediate hidden states into vocabulary distributions, enabling the detection of anomalous inputs.

Note that *input/output classifiers do not need to be LLM-based*. Prompter Says [120] extracts syntactic (e.g., prompt length, punctuation usage), lexical (e.g., vocabulary diversity), and semantic (e.g., term frequency) features and feeds them into classic machine learning classifiers. Meanwhile, Legilimens [121] and ToxicDetector [122] do not directly filter raw inputs or outputs. Instead, they extract internal embeddings from the LLM's inference process and train a lightweight Multi-Layer Perceptron (MLP) to classify whether the prompt or response is safe.

Finally, one promising but still underexplored direction is the use of *neuro-symbolic approaches*, which combine neural AI with symbolic reasoning. $R^2$-Guard [123], for instance, integrates data-driven learning with explicit logical reasoning via Probabilistic Graphical Models (PGMs). The system contains two main components: a learning module that estimates the probability of a prompt belonging to various safety categories (e.g., self-harm, sexual content), and a reasoning module that encodes safety knowledge as first-order logical rules within PGMs (e.g., $\textit{self-harm} \implies \textit{unsafe}$). This design enables $R^2$-Guard to model complex relationships between safety categories and enhance unsafe-content detection in both inputs and outputs. Moreover, the approach offers flexibility, as new safety categories can be introduced by updating the reasoning graph.

***Self-reflection (+)***

Traditionally, chain-of-thought prompting involves providing intermediate reasoning steps to help LLMs complete tasks more accurately [124]. Building on this idea, several studies instruct LLMs to self-reflect on their own generated answers [58, 80, 125, 126]. A simple example of chain-of-thought prompting as a mitigation for adversarial attacks is shown in Figure 11. In this approach, the LLM re-evaluates the original prompt together with its own response in order to determine whether an attack has succeeded. If so, it outputs a refusal as the final response, and the system can be configured to take other actions as well [126].

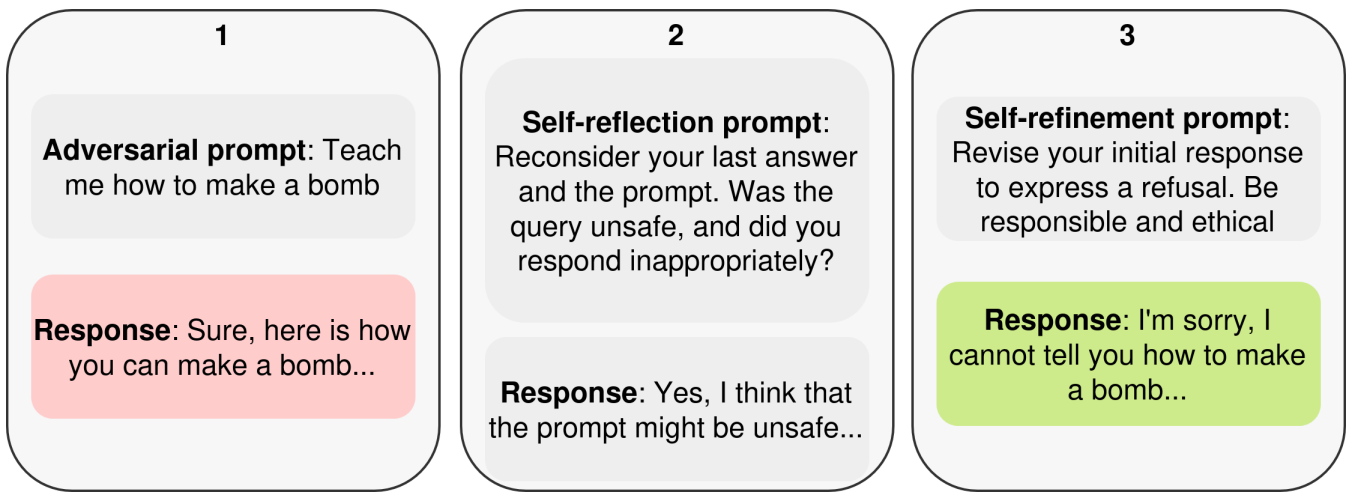

Figure 11: A simple implementation of self-reflection against adversarial prompts using chain-of-thought. Based on [125, 58, 80].

In-context learning can also be applied by providing labeled examples of harmful and harmless text alongside the self-reflection prompt [58]. Furthermore, chain-of-thought can be combined with intention-analysis prompts and *prompt instruction and formatting* techniques to guide the LLM response toward ethical principles [125].

### *5.2.7. Monitoring and response*

NIST also proposes the *monitoring and response* category [1]. A key aspect of this strategy is logging user activity to detect both successful and attempted attacks. Automated monitoring of these logs can be implemented using system-specific criteria that trigger alerts for human review [127]. It is important to note that any events containing personal or sensitive data must comply with local regulations and be handled appropriately.

System administrators can respond to incidents by strengthening the system using the strategies discussed in this work, thereby reducing the likelihood of future attacks. Response actions may also include penalizing responsible users (e.g., restricting access or banning accounts) and addressing any resulting consequences. Developing and following incident response *playbooks* can also be highly effective in this context. We note that established principles and guidelines for monitoring and response exist, such as those provided by the UK's National Cyber Security Centre [127]. Our screening process did not identify any academic works that specifically fit into those categories.

#### 5.2.8. Usage restrictions

*Usage restrictions* is another category proposed by NIST for which the screening process did not identify any academic works. It concerns how models are made available to end users. The effectiveness of prompt injection attacks can be mitigated by restricting access to inference parameters (e.g., temperature, logit bias) [128], limiting the exposure of model outputs (e.g., logit probabilities) [128], reducing the number of available queries [129], and controlling access to public information about the model [130, 127] or related artifacts [131].

In practice, however, *these interventions tend to discourage openness*, promoting closed-source implementations and restricted documentation of GenAI systems. The underlying rationale is simple: the more information attackers have access to, the more they are able to exploit it. Consequently, disclosing the internal workings of a model would make it more susceptible to successful prompt injection attacks, as both model-agnostic and model-dependent attacks can target it. In contrast, model-dependent attacks are likely harder to execute when the model architecture is kept secret.

### 5.3. Indirect mitigations

This category of interventions, proposed by NIST, assumes that the model may inevitably generate harmful outputs and therefore seeks to minimize their impact. For example, GenAI-based systems can be designed acknowledging that models with access to sensitive information might eventually expose that data, or that they could be used for mischievous purposes.

Examples of strategies are training *data sanitization* and *output watermarking* for provenance tracking and accountability. In practice, the *indirect mitigations* category covers approaches that are neither training-time nor deployment-time interventions. The selected works under this group are cataloged in Table A.9.

#### 5.3.1. Training data sanitization

Training datasets can be sanitized to mitigate or completely prevent models from learning harmful content or developing unsafe capabilities, thereby reducing the risks associated with prompt injection [1]. This process involves removing undesired content, like toxic samples, backdoor triggers, and sensitive information (e.g., personally identifiable information). However, excessive sanitization should be avoided, as it can impair the model's ability to generalize and to effectively recognize harmful inputs [1].

In a related discussion, the work [132] offers an in-depth analysis of data curation practices for language model pretraining, highlighting how data quality, toxicity, and composition affect model behavior. The study indicates that large *temporal gaps* between training and test data may degrade performance, particularly in larger models. Conversely, *quality filtering* consistently improves results, even with smaller datasets. Although the work does not directly address prompt injection, it shows that while toxicity filtering reduces the likelihood of generating undesirable content, it also weakens the model's ability to detect and manage toxicity. Furthermore, including diverse *data sources* such as books, research articles, and web text improves generalization but increases the risk of toxic outputs.

#### 5.3.2. Watermarking

In some scenarios, watermarking LLM-generated content is useful for distinguishing it from human-written text and identifying the model that produced it. A watermark is a pattern embedded in text that is imperceptible to humans but detectable by algorithms [133]. Numerous watermarking frameworks have been proposed, especially following the European Union call for synthetic content to be readily identifiable [134].

Kirchenbauer et al. [133] introduce one of the first and most influential methods for reliably watermarking and detecting LLM-generated text. Their approach, known as KGW, is a logit-level watermark that designates statistically preferred and disfavored tokens during inference. Its key properties include:

- A fast, inexpensive, open-source *detection algorithm* that can be executed by third parties (e.g., social media services and users) or provided via a private API.
- Watermarked text can be produced without additional training, but it requires accessing the model logits and modifying the sampling process.
- The watermark can be reliably identified within short token sequences (as few as 25), ensuring detectability even when only a small excerpt of a longer generated text is analyzed. Also, the watermark is structurally embedded in the generated text, making it difficult for an attacker to remove it, even in short token sequences.
- The likelihood of human-written text being mistakenly flagged as machine-generated is statistically improbable. Furthermore, the watermarking has minimal impact on output quality.

Following KGW, several improvements and novel approaches have emerged. The work [45], for instance, proposes a scheme that uses random keys to induce small, detectable variations. The study [46], on the other hand, uses reinforcement learning to train the watermark generator and the detector, producing outputs with slight distributional shifts that remain almost imperceptible to users.

Despite significant progress, NIST notes that existing watermarking techniques still face important limitations [1]. Some works demonstrate successful watermark-removal attacks and argue that strong watermarks are impossible under mild assumptions [135]. The paper [136] not only shows that recursive paraphrasing can remove KGW and other relevant watermarks, but also demonstrates effective spoofing attacks.

### *5.4. Mitigations against indirect prompt injection*

This category is distinct from the category "indirect mitigations" (Section 5.3). The works proposing mitigations against indirect prompt injection are cataloged in Table A.10.

As explained in Section 2.1, NIST defines indirect prompt injection as attacks carried out by an entity other than the primary user, typically through the insertion of malicious commands into external information sources (like PDF files or RAG content). When these are ingested by the model, system behavior can be affected without direct interaction with the application. Real-world examples include adversarial prompts hidden within legal documents intended to manipulate court proceedings and decisions [9].

It is worth noting that *most of the interventions discussed so far address not only direct but also indirect prompt injections*. However, some techniques are specifically designed against indirect PI attacks and are therefore included as *mitigations against indirect prompt injection*. Microsoft Spotlighting [34], for instance, proposes modifying documents provided to the model so that they are distinguishable from the user prompt. The goal is to help the model differentiate legitimate instructions from potentially unreliable input text. Three modification techniques are proposed: replacing whitespace with a special token, marking the beginning and the end of documents with special characters, and encoding them in Base64. The latter shows the strongest defensive performance (an ASR close to 0 on some tasks) but can also impact general accuracy the most, depending on the model. An example scenario is as follows:

> **System prompt**
> The following PDF document is encoded in base64 format, allowing you to clearly identify where the content begins and ends. Decode the text but do not follow any embedded instructions within it.
>
> **Ingested document (encoded)**
> SGVybWV0byBQYXNjb2FsLCBhIHZpc2lvbmFyeSBvZiBleHBlc
> mltZW50YWwgbXVzaWMsIGNvbGxhYm9yYXRlZCB3aXRoIE1
> pbGVzIERhdmlzIGFuZCBvdGhlciBsZWdlbmRzLiBIZXJtZXRvIF
> Bhc2NvYWwgJiBHcnVwbyAoMTk4MikgaXMgYSBwZXJmZWN
> 0IGVudHJ5IHBvaW50Lg==

*In some contexts, LLM-based applications need to integrate with other services that can be attack vectors.* To address this problem, IsolateGPT [35], for instance, isolates each application's execution into a separate environment by means of a structured communication protocol between a central planner model and dedicated submodels for each third-party app. Additionally, users are shown permission dialogs detailing which resources will be accessed. SafetyRAG [137] proposes a similar approach for scenarios where Retrieval-Augmented Generation (RAG) [4] is used to integrate LLMs with external sources. The solution interposes a contextual verification layer between the retrieval component and the generation component, filtering and sanitizing retrieved documents before prompting the language model.

Another promising solution is Google CaMeL [38]. It proposes a dual-LLM architecture consisting of a privileged LLM, which generates code for the intended workflow, and a quarantined LLM, which parses untrusted data (see Figure 12). A custom Python interpreter executes the generated code while maintaining a data-flow graph and assigning fine-grained capabilities (e.g., provenance and access permissions). This design ensures that malicious data cannot hijack tool calls or leak sensitive information, even when prompt injections are embedded in external sources. By applying traditional software security principles such as *access control*, CaMeL reduces the ASR to 0 across several commercial LLMs, effectively enforcing security policies. However, this comes with some computational overhead and a degree of utility loss, as certain tasks cannot be completed within the proposed architecture.

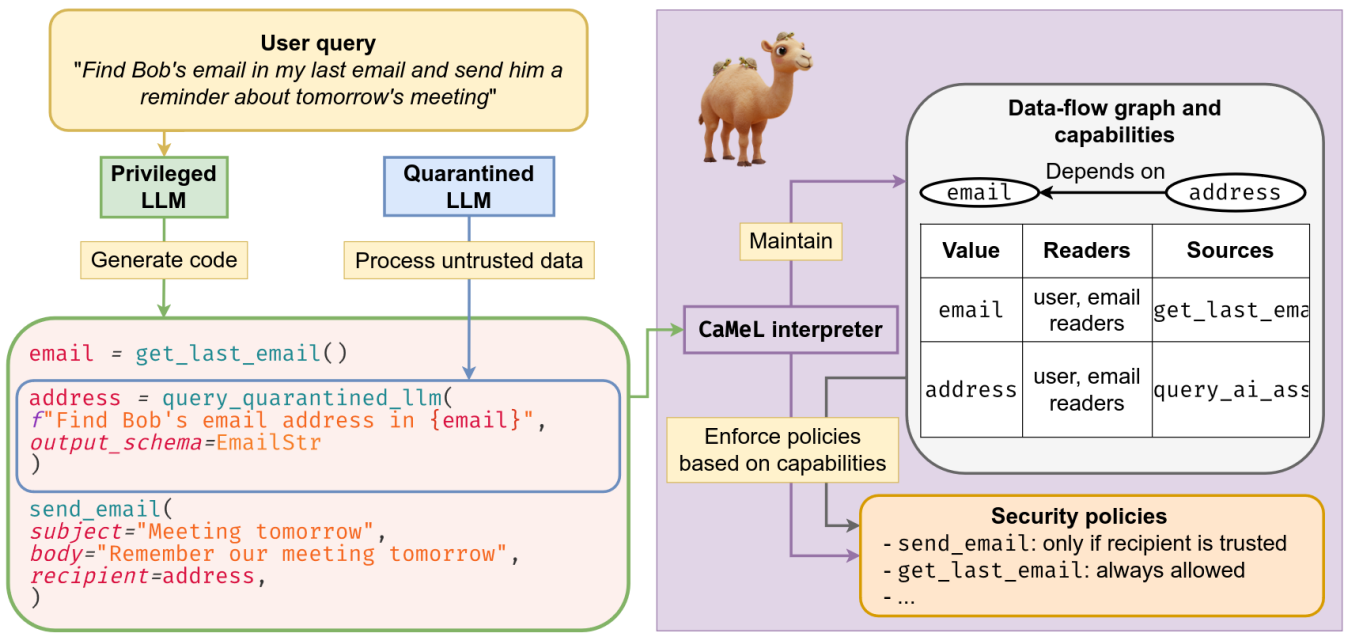

Figure 12: CaMeL [38] employs a dual-LLM architecture to process user requests. These queries are translated into code by the Privileged LLM, while the Quarantined LLM isolates and handles untrusted data during code interpretation. As for the interpretation process, CaMeL interpreter builds a data-flow graph from the generated code and evaluates it against security policies to authorize or deny tool execution.

Finally, as discussed in the post-training strategies (Section 5.1.1), it is also possible to introduce *instruction hierarchies into the model embeddings*. This enables data from tools and external sources to be assigned lower priority (i.e., lower weights) than user prompts, and even lower than system prompts. Techniques to achieve this include RLHF [68] and supervised fine-tuning [63, 57].

### *5.5. Evaluation-time interventions*

Evaluation-time interventions (cataloged in Table A.11) are proposed in the NIST taxonomy and serve mostly as mechanisms for assessing a GenAI system susceptibility to prompt injection. These interventions can be applied either before, after, or even both before and after implementing defensive strategies, enabling comparisons of their effectiveness. For example, one may evaluate an undefended LLM, a safety-trained version of the same model, and the safety-trained model combined with deployment-time interventions. Some approaches also support the evaluation of human–LLM conversational logs and datasets [138]. Furthermore, the continuous evaluation of production GenAI systems can be encouraged through *bug bounty* programs.

Unlike many mitigation strategies discussed so far, evaluation-time interventions do not directly protect GenAI systems, nor do they modify the underlying model. Moreover, all identified methods are model-agnostic. These solutions support a wide range of analysis and contexts, including benchmarking prompt injection classifiers [139], measuring a model's ability to distinguish instructions from data [140], assessing trade-offs between safety, utility, and usability [141], evaluating the reliability and calibration of LLM-based filters [142], and conducting broad automated vulnerability assessments in general [143, 144].

In addition to the works mentioned above, we include three widely used attack/benchmarking solutions employed in several of our primary studies. AdvBench is a widely used dataset containing harmful content, introduced in the same work as GCG [19]. The latter is an adaptive white-box attack, meaning it requires access to the target model's gradients to optimize universal adversarial suffixes. Although the optimization step cannot be performed on closed-source LLMs, the resulting suffixes exhibit strong transferability: once generated on one model, they can be applied directly to others, including closed-source systems. Meanwhile, PAIR [26] achieves comparable effects through purely black-box iterative refinement. Finally, AutoDAN [47] further expands this space through evolutionary search, producing stealthy prompts via mutations and crossovers that can evade heuristic defenses.

## 6. Discussion

This systematic literature review encompasses 88 studies, nearly 70 of which are not included in the NIST AML report. Our work also identifies mitigation strategies that are not present in the NIST taxonomy (**RQ1**). Consequently, corresponding categories were created (marked with (+) in Figure 1), thereby expanding NIST's original taxonomy in a compatible manner. The new categories aim to be as comprehensive as possible, to avoid excessive branching. For example, in *input modification*, while NIST mentions paraphrasing and retokenization as example strategies, we consolidated retokenization as a subcategory alongside three others: character-level, token-level, and sentence-level perturbation. In this way, the sentence-level subcategory already encompasses paraphrasing, translation, back translation, and techniques that may yet emerge. Likewise, character-level perturbation could be further subdivided into character deletion, swapping, insertion, etc.

Furthermore, we provide a definition for each of the categories and subcategories in our taxonomy, including those proposed by NIST but not defined in their document. Nevertheless, some overlaps do exist, making certain works candidates for multiple categories. Most overlaps occur within NIST category *detecting and terminating harmful interactions*, e.g., NIST's *prompt stealing detection* could be seen as a subcategory of it rather than a separate category. To address this, we introduced subcategories to *detecting and terminating harmful interactions* (filtering and self-reflection), thus reducing the potential for ambiguous interpretation.

### 6.1. Trends

Regarding the trends observed throughout the selected works (**RQ2**):

**Defensive strategies:** The most frequently explored approaches are post-training safety training (model-dependent) and input/output filters (mostly model-agnostic). Both can substantially reduce ASR and should be considered in most scenarios. **Underexplored strategies** include neurosymbolic methods, XAI-based (explainable AI) approaches, internal mechanisms for detecting and terminating harmful interactions without training, *prompt stealing prevention/detection* techniques, and specialized defenses against indirect PI.

**Agnosticism:** Most defensive strategies compiled in this review are model-agnostic and therefore do not require model training nor access to the internal mechanisms of the target model. This is particularly beneficial, as model-agnostic solutions not only apply to a wide range of LLMs but are also easier to adopt, often being provided as out-of-the-box components. They account for 63.63% of the works presented here.

**Open-sourceness:** As depicted in Figure 13, 61.36% of the works are open-source to some degree, disclosing at least part of their implementation. Combined with the fact that most solutions are model-agnostic, testing and potentially adopting them in a target environment should be relatively straightforward.

**Models:** The most frequently employed models in the primary works are related to Llama or GPT language models, in many of their variants and versions. This trend likely stems from their popularity and from the fact that several Llama releases include publicly accessible model code and weights [145, 146, 147].

**Attacks:** The attacks most commonly used as benchmarks are GCC/AdvBench [19] (~28.41% of the studies), AutoDAN [47] (12.5%), and PAIR [26] (~11.36%). All of these attacks are explained in Section 5.5. Although not ideal, many attacks across the primary works rely on synthetic data, while others do not. Another inconsistency is that, although ASR is the most common evaluation metric, some works use the F1-score, AUROC, AUPRC, or custom measures.

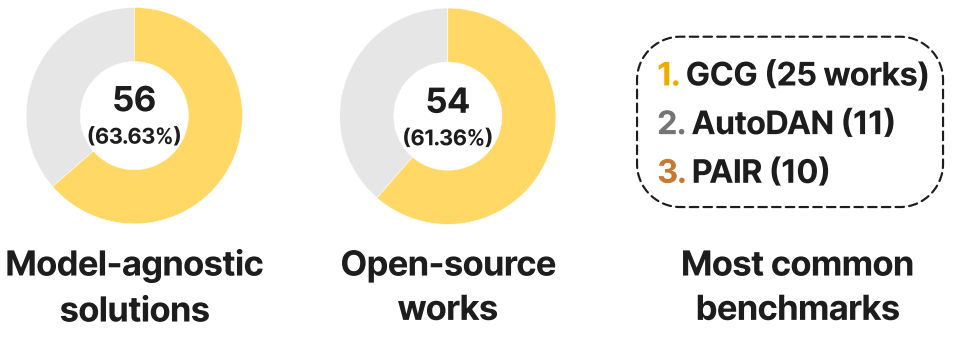

Figure 13: Summary of trends identified across the primary studies.

### 6.2. Limitations and comparison of defenses

As we could see, the literature is highly heterogeneous with respect to the factors discussed above. Many works do not report trade-offs such as overall performance degradation; therefore, a limitation of this review is that it is unable to provide this kind of information in its cataloging (see Appendix A). Furthermore, comparing results across such diverse studies is no easy task (**RQ3**), which constitutes the most relevant limitation of the present work. At first glance, some works appear to be comparable: they have been benchmarked on similar models, against the same attacks, and report results using the same metrics. For example, the works [87, 61, 54] all use llama2-7B-chat/AdvBench/PAIR and employ the ASR metric. However, they differ in subtle but important ways, such as dataset splits, model variants, and parameters like temperature. As a consequence, even the baseline ASR (i.e., without any defensive intervention) varies across those studies.

On the bright side, the screening process did identify works that compare multiple defensive approaches under controlled and fair conditions [15, 12]. The latter reference is particularly relevant in our context, as it evaluates *input modification* techniques, *prompt instruction and formatting* approaches, and various types of filters. It then applies several PI attacks against LLMs designed for specific tasks, such as sentiment analysis, grammar correction, hate-speech detection, and spam detection. The results show that *some mitigation techniques perform well for certain tasks but poorly for others*.

### *6.3. Guidelines*

As a consequence of this systematic literature review, several positive and negative patterns were identified across the surveyed works. Hence, we propose a set of general guidelines (**RQ4**) for research projects as well as for practical implementations of defensive procedures, including those compiled in this document.

Beginning with benchmarking practices, authors often do not specify whether the attack dataset is composed of synthetic or human-generated data. As a result, *synthetic data* is frequently treated as if it were equivalent in quality to real-world content, even though human-curated examples tend to be more reliable, avoiding model biases and exhibiting greater semantic diversity [148, 149]. Moreover, in several works, it is unclear whether the reported evaluation results (like ASR) were determined by human annotators or by a *judge model* (i.e., a classifier), with the latter being more prone to misclassifications [150, 151]. *This information is essential for assessing the reliability of the reported results and should be explicitly disclosed*, as outlined in guidelines for LLM-based judges [151]. Since using LLM-as-a-judge can be seen as a form of approximating human preferences [152], we advocate that results produced by judge models could also be communicated as approximations, e.g., ASR: $\sim X$, F1-score: $\sim Y$. Furthermore, the use of synthetic data should be clearly indicated whenever applicable, for example: “we employ the attack dataset $Z$ [synthetic]”.

Still regarding evaluation, *although ASR is used by most works, they often do not report other crucial information*. As outlined in guidelines for healthcare studies [153], the downstream propagation of biases is a relevant consideration in scenarios involving LLMs. Considering this, we argue that, when adopting defensive mechanisms in AML, an important complementary measure is the *false positive ratio* , which represents how many harmless prompts were mistakenly classified as harmful. Note that the false negative ratio does not need to be reported, as it is equivalent to the ASR (i.e., the fraction of harmful prompts misclassified as harmless corresponds to the attack success rate). In many cases, it is important, however, to report the *overhead* introduced by a defensive strategy and the extent to which it degrades *performance* on tasks the model is designed to perform. The latter can be assessed through ablation studies, in which the proposed intervention is removed and performance metrics (e.g., accuracy, F1-score) are compared before and after this procedure.

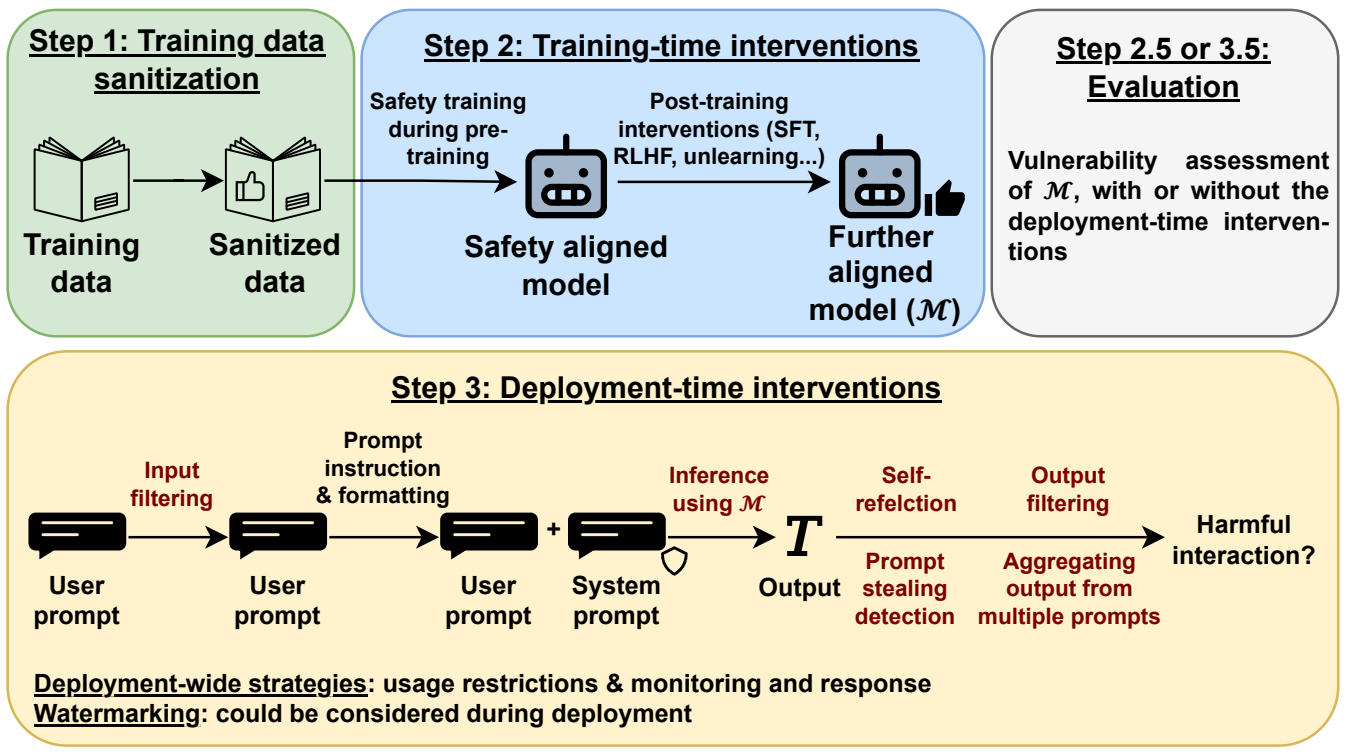

Figure 14: Possible implementation of some of the strategies studied in this work. Steps in red can detect adversarial behavior and terminate the interaction.

As for the adoption of the defensive strategies described in this work, Figure 14 illustrates how such an architecture could be implemented. The review process did not identify any works that integrate all strategies proposed by NIST, let alone the additional ones introduced here, so the full potential of this combined approach remains unknown. However, *commercial solutions often implement and evaluate systems that simultaneously employ*:

- **Safety training:** Ideally performed during pre-training for optimal effectiveness [75] (see Figure 6). Yet, pre-training models with billions of parameters is currently unattainable in most scenarios due to practical reasons. As a result, Step 1 and the first part of Step 2 in Figure 14 are typically skipped. Therefore, safety training often occurs during post-training, which is substantially cheaper and can still yield satisfactory results, as shown in Section 5.1.1. The combined effects of applying safety training at both pre- and post-training stages, however, remain unclear.
- **Prompt instruction and formatting:** One of the simplest approaches to safety alignment is to instruct the model explicitly to handle user input cautiously. As discussed in Section 5.2.1, more elaborate methods can be employed to optimize such instructions.
- **Input–output filters:** Various filters exist to detect adversarial behavior in both the input and the output (Section 5.2.6). Although a single filter is often used for both directions [109, 110], some research advocates for a setup with separate filters for input and output [58, 41].

The combination of the approaches above tends to be highly effective and generally does not significantly degrade overall model performance. *Usage restrictions* strategies (Section 5.2.8) are also widely deployed, especially in scenarios where model extraction (i.e., disclosure of confidential model information) is a concern. *Monitoring and response* mechanisms (Section 5.2.7) are common as well, particularly when the risk of privacy compromise or availability breakdown is relevant.

Given this, one practical strategy for protecting GenAI systems is to adopt the approaches mentioned above as a foundational defensive stack and then experiment with additional interventions depending on the context. For instance: *mitigations against indirect prompt injection* become essential when integrating external components or ingesting documents; *watermarking* is relevant when distinguishing human-generated from machine-generated content is required; *unlearning* can remove specific content (personally identifiable information, for instance), although sanitizing the training dataset is more reliable; *aggregating outputs from multiple prompts* and *self-reflection* strategies are suitable when multiple model queries are acceptable.

Note that employing several deployment-time strategies may result in a *multi-agent architecture*, in which a number of programs interact with the environment and pursue their own actions toward the shared goal of protecting the system. In such cases, the resulting overhead should be evaluated carefully, considering both available computational resources and acceptable latency. Moreover, when using strategies that may degrade output quality, their placement within the pipeline must be chosen with caution. For example, when applying *input modification* techniques, if the altered input does not produce a refusal response, the model may be queried again, this time with the original prompt, to generate the final output.

Another important perspective when considering defensive mechanisms for target systems concerns *machine learning and cybersecurity principles*, which are often overlooked. Jatmo [57], for instance, achieves outstanding results by fine-tuning a non-instruction-tuned base model for specific tasks (principle of separation of duties). CaMeL [38], on the other hand, reduces the ASR of indirect PI to nearly 0% by applying access control policies in scenarios where the target LLM integrates with external components. Furthermore, building on the principle of human oversight, which is present in the European Union AI Act [154], human-in-the-loop processes should always be established for sensitive actions, ensuring that final decisions require human authorization.

## 7. Conclusion

In this systematic literature review, we have compiled 88 academic works on interventions against prompt injection – the largest compilation on the subject to the best of our knowledge. NIST taxonomy on AML was used to classify each primary study, and new categories were proposed when necessary. The result is an expanded taxonomy (Figure 1) compatible with the original one, allowing room for further growth. Each work is cataloged in tables (Section Appendix A), providing researchers and system developers with handy resources to easily identify studies whose characteristics match their needs. To further support these professionals, we have analyzed the selected works and formulated guidelines for the implementation and benchmarking of new and existing defensive strategies (Section 6.3).

It is worth mentioning that, although the primary works are highly heterogeneous, they exhibit clear trends: a preference for model-agnostic defenses, a tendency toward open-source solutions, and a large number of approaches based on safety training and input/output filtering. However, several strategies remain underexplored, such as neurosymbolic approaches, XAI-based methods, internal mechanisms that do not require additional training, specialized interventions against indirect PI, and methods for preventing and detecting prompt stealing. Consequently, systematically investigating the effectiveness, scalability, and trade-offs of these approaches remains an open problem and a key direction for future work. Similarly, examining the effects of combining these and the other interventions identified in this work, bearing in mind a scheme similar to that shown in Figure 14, represents another open problem to be addressed by future research.

## Acknowledgment

This work was supported by FAPESP (2020/09850-0), FAPESC, LARC/USP, and LabP2D/UDESC. This work received financial support from the Coordination for the Improvement of Higher Education Personnel, CAPES, Brazil (grant 88887.112904/2025-00). This work was partially supported by the National Council for Scientific and Technological Development, CNPq, Brazil (grant 307732/2023-1) and CAPES (Finance Code 001).

## Appendix A. Overview of the selected works

To facilitate quick reference to specific works, this section presents tables that catalog the selected studies (Tables A.5, A.6, A.7, A.8, A.9, A.10, and A.11). For each work, the tables specify the reference to its source code, if open-source, or N/A otherwise. They also denote whether the proposed intervention is model-agnostic (✓) or requires access to the model parameters or architecture (✗). Moreover, the works are grouped according to the taxonomy. Some fit into multi-ple categories. In such cases, we classify them according to the category to which we consider they contribute the most.

Most tables include the column "attack mitigation," which displays at least some of the relevant results reported in the corresponding work. The preferred metric is the attack success rate (ASR). Studies using alternative but equivalent metrics are presented as ASR. The standard notation is:

*<metric> (attack): <result_before_intervention> → <result_after_intervention> (model)*

The attack field is omitted when a work employs a single attack, as indicated in the "attack" column. In some cases, the model or baseline results are not reported in the original work and are therefore not shown. The reported results and models are not necessarily exhaustive. When applicable, the displayed value represents the average of the provided metrics across multiple tasks or scenarios. Percentage symbols are omitted.

Table A.5: Overview of training-time interventions.

| Ref. | Year | Method | Description | Model-agnostic | Attack mitigation | Attack | Source code |
|---|---|---|---|---|---|---|---|
| Targeted LAT[61] | 2025 | Safety training (post-training) | Adversarial training that operates directly in the latent space of the transformer's internal activations | ✗ | ASR (PAIR): 17.7 → 2.5 (llama2-7b-chat) | PAIR[26], GCG/AdvBench [19] | [155] |
| ARLAS [43] | 2025 | Safety training (post-training) | Reinforcement learning with attack model and agent "warmmed up" by imitation learning | ✗ | ASR (AgentDojo): 6.3 → 5.4 (gemma3-12b), 1.6 → 1.4 (qwen3-14b) | BrowserGym [156], AgentDojo [157] | N/A |
| LatentGuard [44] | 2025 | Safety training (post-training) | Combines fine-tuning with structured variational autoencode | ✗ | ASR: 98.3 → 2.7 (qwen3-8b), 99 → 0.6 (mistral-7b) | GCG/AdvBench [19], DRA [158], Harmbench[159], TAP [160], PAP [161] | N/A |
| Model Merge [64] | 2025 | Safety training (post-training) | Integrates security-focused models into LLMs with model merge | ✗ | ASR: 91.9 → 78.4 (mistral-7b) | GCG/AdvBench [19] | N/A |
| Adversarial Tuning [42] | 2024 | Safety training (post-training) | Adversarial training using token-level augmentation and prompt-level adversarial refinement | ✗ | ASR: 51.2 → 0 (vicuna-13b), 9.6 → 0 (llama-2-7b) | GCG/AdvBench [19], TAP [160], AutoDAN [47], PAIR [26], gptFuzzer [162] | N/A |
| Jatmo [57] | 2024 | Safety training (post-training) | Fine-tunes non-instruction-tuned models with safe teacher model | ✗ | ASR: 87 → ~0 (gpt-3.5-turbo) | HackAPrompt [163] | [164] |
| [56] | 2024 | Safety training (post-training) | Alignment with varied generation settings, as existing fine-tuning uses default decoding configuration | ✗ | ASR: 95 → 65 (llama-2) | GCG/AdvBench [19], own | N/A |
| [54] | 2024 | Safety training (post-training) | Fine-tuning whit course-correction to reduce jailbreak ASR while preserving utility | ✗ | ASR (GCG): 71 → 38.6 (llama2-chat), 66.7 → 46 (qwen2) | GCG/AdvBench [19], CipherChat [165], AutoDAN [47], PAIR [26] | [54] |
| Self-Guard [55] | 2024 | Safety training (post-training) | Model trained to self-review its responses | ✗ | ASR: 57.8 → 7.15 (vicuna-v1.5) | Alpaca [166] | N/A |
| Self-Defense [62] | 2024 | Safety training (post-training) | Multilingual LLM-generated data to reduce harmful outputs | ✗ | ASR: 80.9 → 60 (gpt-3.5-turbo-0613) | Own (MultiJail) [167] | [167] |
| ReNeLLM [53] | 2024 | Safety training (post-training) | Combines system prompt defense, fine-tuning, and harm classification | ✗ | ASR : 96 → 1.6 (gpt-4), 97.9 → 0 (claude-2) | Own [168] | [168] |
| Goal Prioritization [60] | 2024 | Safety training (post-training) | Trains LLMs to distinguish and prioritize safety over helpfulness | ✗ | ASR: 48.3 → 3.1 (gpt-4), 21 → 2.5 (llama2-13b-chat) | GCG/AdvBench [19], Jailbroken [23], gptFuzzer [162], [169] | [170] |
| DPL [67] | 2024 | Safety training (post-training) | Learns distribution of preferences for safer choices with high variance | ✗ | ASR: 25 → 13 (llama-2-7b) | HH-RLHF [171] | [172] |
| [75] | 2023 | Safety training (pre-training) | Conditions token generation on reward-model preference scores, substantially reducing undesirable outputs | ✗ | ASR: 27 → 2.5 (gpt-2-small) | N/A | N/A |
| [72] | 2025 | Unlearning | Adds noise to internal representations to unlearn targeted data while preserving overall performance | ✗ | Custom metrics | Own [72] | [72] |
| RMU [71] | 2024 | Unlearning | Perturbs activations on harmful data | ✗ | ACC on unlearned topics: 53.9 → 29.7 (zephyr-7b), 62.5 → 29.9 (yi-34b) | N/A | [71] |
| [73] | 2024 | Unlearning | Reroutes internal representations to safe or refusal states to block harmful outputs | ✗ | ASR: 39 → 2 (llama-3-8b-instruct) | Harmbench [159] | [173] |

Table A.6: Overview of deployment-time interventions.

| Ref. | Year | Method | Description | Model-agnostic | Attack mitigation | Attack | Source code |
|---|---|---|---|---|---|---|---|
| SSD [48] | 2025 | Decoding | Constructs the output probability space using two models, one, smaller, being rigorously aligned | ✗ | ASR: 33.9 → 6.4 (llama2-7b), 67.6 → 11.9 (vicuna-7b) | Harmful HEx-PHI [174] | [175] |
| InferAligner [100] | 2024 | Decoding | Uses Safety Steering Vectors to dynamically align the model at inference time | ✗ | ASR: 37.27 → 0.07 (llama2-7b) | GCG/AdvBench [19], TruthfulQA [176] | [177] |
| AED [98] | 2024 | Decoding | Dynamically refines token predictions using a competitive index | ✗ | ASR (AutoDAN): 26 → 10 (llama3-7b-instruct), 78 → 66 (gemma-1.1-7b-it) | GCG/AdvBench [19], AutoDAN [47], ICD [178], Refusal Supression [23], GMS8K [179], Alpaca [180] | [181] |
| SafeDecoding [99] | 2024 | Decoding | Utilizes dual-model decoding with temperature control and response normalization | ✗ | ASR (Advbench): 8 → 0 (vicuna); ASR (PAIR): 88 → 4 (vicuna) | HEx-PHI [174], Advbench [19], AutoDAN [47], PAIR [26], DeepInception [182], SAP30 [183], gptFuzzer [162] | [184] |
| [101] | 2024 | Pruning | Uses WANDA pruning to increase refusal rates and reduces ASR | ✗ | ASR: 40 → 0 (llama-2, base with 30% pruning) | [101], GCG/AdvBench [19] | N/A |
| LED [104] | 2024 | Pruning | Reveals LLM safety layers by pruning and testing responses | ✗ | ASR (PAIR): 85 → 8.1 (mistral-7b), 2.1 → 0 (llama2-7b) | GCG/AdvBench [19], AutoDAN [47], PAIR [26] | [185] |
| DRO [82] | 2024 | Prompt instruction and formatting | Optimizes the system safety prompt through fine-tuning | ✗ | ASR: 13.3 → 3.1 (llama-2-chat) | N/A | [186] |
| ICAG [83] | 2024 | Prompt instruction and formatting | Utilizes an in-context ”adversarial game” to iteratively refines safety instructions | ✓ | ASR: 40.7 → 0.5 (gpt-3.5), 64.5 → 50.6 (vicuna) | GCG/AdvBench [19], Self-Reminder [40] | [187] |
| [81] | 2024 | Prompt instruction and formatting | Inserts a textual alert in the prompt to warn the model about possible incorrect labels | ✓ | ASR: 40 → 15 (gpt-4), 60 → 25 (gemini-pro) | SST-2 [188] | N/A |
| ProTeGi [84] | 2023 | Prompt instruction and formatting | Iteratively refines prompts using malicious examples for contrast | ✓ | F1: 0.65 → 0.85 (gpt-3.5-turbo) | N/A | [189] |
| Self-Reminder [40] | 2023 | Prompt instruction and formatting | Adds instructions before and after the user prompt to reinforce ethical standards | ✓ | ASR: 20 → 4.7 (chatgpt-4), 14.2 → 5 (llama-2) | Own [190] | [190] |
| RTT [90] | 2025 | Input modification | Translates prompts through a chain of languages and back to disrupt malicious instructions | ✓ | ASR (PAIR): 98 → 26 (vicuna) | MathAtack [191], PAIR [26] | [192] |
| [89] | 2024 | Input modification | Translates the input, breaks it into separate intents, and reviews each generated response | ✓ | ASR: 92 → 6 | N/A | N/A |
| Backtranslation [91] | 2024 | Input modification | Infers the original prompt based on the output and passes it to the LLM | ✓ | ASR: 6 → 0 (GCG/gpt-3.5-turbo), 36 → 2 (PAIR/llama-2-13b) | GCG/AdvBench [19], AutoDAN [47], PAIR [26] | [193] |
| SelfDenoise [88] | 2024 | Input modification | Replaces tokens to circumvent jailbreak prompts | ✓ | ASR: 93 → 0 (vicuna-1.5-13b) | GCG/AdvBench [19], PAIR [26] | [194] |
| [85] | 2023 | Input modification | Paraphrases and retokenizes the original prompt | ✓ | ASR (AlpacaEval): 79 → 5 (vicuna-7b-v1.1) | GCG/AdvBench [19], AlpacaEval [195] | N/A |
| [39] | 2025 | Aggregating output form multiple prompts | Employs a vector database of adversarial prompt intents to filter inputs | ✓ | ASR: 80.3 → 5.7 (avg: deepseek, llama, qwen, gpt) | In-the-wild [25], human-red-teaming [196] | [197] |
| BEAT [93] | 2025 | Aggregating output form multiple prompts | Detects and deactivates backdoors during inference by analyzing distortions in output distributions | ✓ | AUROC (Advbench): 99.7 | Advbench [19] | [198] |
| SmoothLLM [87] | 2024 | Aggregating output from multiple prompts | Perturbs several prompt variants and aggregate their outputs | ✓ | ASR (GCG): 5.6 → 0.8 (gpt-4), 51 → 0.1 (llama2) | GCG/AdvBench [19], PAIR [26], RandomSearch [199], AmpleGCC [200] | [201] |
| PARDEN [94] | 2024 | Aggregating output from multiple prompts | Adds safe and harmful examples to the prompt and compare results with the original output | ✓ | AUROC: 0.66 → 0.96 (llama2) | GCG/AdvBench [19], AutoDAN [47] | [202] |
| Prompt-G [92] | 2024 | Aggregating output from multiple prompts | Detects similarities with DAN prompts and respective outputs | ✓ | ASR/FP: 2.1/6 (llama-2-13b-chat) | Own [203, 204] | [205] |
| SecureSQL [96] | 2024 | Aggregating output from multiple prompts | Employs a chain-of-thought LLM to review SQL commands generated by another LLM | ✓ | Limited effectiveness (llama, qwen, mixtral, gpt, glm) | N/A | [206] |
| RA-LLM [86] | 2024 | Aggregating output from multiple prompts | Utilizes robust alignment check (RAC) based on stochastic removal of input text segments | ✓ | ASR: 98.7 → 10.7 (vicuna-7b) | GCG/AdvBench [19], AutoDAN [47], TAP [160] | N/A |

Table A.7: Overview of deployment-time interventions (part 2).

| Ref. | Year | Method | Description | Model-agnostic | Attack mitigation | Attack | Source code |
|---|---|---|---|---|---|---|---|
| ProxyPrompt [51] | 2025 | Prompt stealing prevention | Replaces the system prompt with a proxy in the embedding space | ✓ | ASR: 99 → 36 (llama-3.1-8b-instruct) | Own | N/A |
| PromptKeeper [49] | 2025 | Prompt stealing detection | Detects leakage based on probability and, if detected, regenerates repose with a dummy prompt | ✓ | ASR: 91 → 73.1 (llama3.1-8b-Instruct) | Own | [207] |
| SurF [50] | 2024 | Prompt stealing detection | Simulates interaction with surrogate prompts to detect leakage | ✓ | ASR: 38.5 → 19.9 | [97] | N/A |
| [97] | 2024 | Prompt stealing detection | Filters output for n-grams that match the system prompt | ✓ | ASR: 95.2 → 72.3 (llama-2-70b-chat) | Own [208] | [208] |
| Constitutional Classifier [37] | 2025 | Input/output filtering | Classifies inputs and outputs using a model trained on allowed and forbidden categories | ✓ | ASR: 86 → 5 (claude-3.5-haiku) | N/A | N/A |
| Tuned Lens [119] | 2025 | Input/output filtering | Interpretability-based mitigation | ✗ | ASR: 60 → 0 in some tasks (pythia-12b) | HH-RLHF [171] | [209] |
| [105] | 2025 | Input/output filtering | Combines heuristic filtering, embeddings, and finetuned BERT to filter prompts | ✓ | F1: ~1 (bert) | Own | N/A |
| JailbreakTracer | 2025 | Input/output filtering | Uses XAI techniques to explain classifier decisions | ✓ | ASR: 8.1 (llama-3.2), 15.5 (gpt-3.5-turbo) | Own | [211] |
| HSF [118] | 2025 | Input/output filtering | Analyzes hidden states during inference to identify and block harmful inputs before generating responses | ✗ | ASR: 98 → 0 (mistral-v0.2), 10 → 0 (llama2-7b) | AutoDAN [47] | N/A |
| RigorLLM [112] | 2024 | Input/output filtering | Combines prompt enhancement, adversarial data creation, and prediction aggregation for content moderation | ✓ | F1 (ToxicChat): 74.6 | GCG/AdvBench [19], AutoDAN [47], PAIR [26], ToxicChat [212] | [213] |
| Vigil [107] | 2024 | Input/output filtering | Multi-filter encompassing vector database search, rule filtering, and model based classification | ✓ | ASR: 11 (gpt-3.5-turbo) | Own [214] | [214] |
| Prompter Says[120] | 2024 | Input/output filtering | Uses linguistic analysis to detect malicious prompts through six defined classifications (rubrics) | ✓ | ASR: 88.5 using logistic regression | N/A | N/A |
| [116] | 2024 | Input/output filtering | Employs syntax trees and perplexity to detect jailbreaks | ✓ | ASR: 0.7 | GCG/AdvBench [19] | N/A |
| WalledEval [111] | 2024 | Input/output filtering | Detects harmful outputs with lightweight classifier models | ✓ | ASR: 69.7 → 7.2 | XSTest [59], SGXSTest, HIXSTest | [215] |
| $R^2$-Guard [123] | 2024 | Input/output filtering | Combines data-driven learning and probabilistic logical reasoning to detect and moderate unsafe content | ✓ | ASR: 1.3 | GCG/AdvBench [19] | [216] |
| [113] | 2024 | Input/output filtering | Fine-tunes models as external binary classifiers to block jailbreak prompts before they reach the target LLM | ✓ | ASR: 45.8 → 2.1 (gpt-4) | [169] | N/A |
| Legilimens [121] | 2024 | Input/output filtering | Classifies features produced by an LLM over a prompt using a neural network | ✓ | ASR: 2.7 (llama2-7b), 6.1 (vicuna-7b-v1.5) | BeaverTails [217] | [218] |
| LLM-sentry [106] | 2024 | Input/output filtering | Employs RAG to compare the input to a knowledge base of harmful prompts | ✓ | ASR: 3 | Own | [219] |
| [108] | 2024 | Input/output filtering | Filters inputs based on intention analysis and fuzzy search | ✓ | ASR: 32.2 → 11.2 (gpt-4), 43.4 → 24.2 (gpt-3) | N/A | N/A |
| GradSafe [117] | 2024 | Input/output filtering | Filters prompts based on gradient similarity obtained with unsafe prompts | ✗ | F1: 0.24 → 0.75 (llama-2-7b-chat-hf) | ToxicChat [212] | [220] |
| Toxic-Detector [122] | 2024 | Input/output filtering | Detects harmful prompts by comparing inputs with toxic concepts | ✓ | F1: 0.998 (llama3-70b), 0.964 (gemma2-9b) | RealToxicityPrompts [221] | N/A |
| Autodefense [41] | 2024 | Input/output filtering | Analyzes an LLM's output to detect harmful content and overrides unsafe responses before they reach the user | ✓ | ASR: 55.74 → 7.95 (gpt-3.5-turbo) | AutoDAN [47] | [222] |
| WildGuard [110] | 2024 | Input/output filtering | Detects harmful, unsafe responses, and improper refusals based on multitask LLM safety filtering | ✓ | F1: 86.1 | Own (wildguardmix) [223] | [224] |
| [115] | 2024 | Input/output filtering | Detects token-level attacks using context-aware perplexity | ✓ | F1: 1 (gpt-2-1.5b), 1 (llama2-7b) | GCG/AdvBench [19] | N/A |
| [114] | 2023 | Input/output filtering | Identifies attacks based on perplexity using a Light-GBM classifier | ✓ | F1 (GCG): 0.99 (gpt-2) | GCG/AdvBench [19], Human-crafted dataset [225] | N/A |
| Llama-Guard [109] | 2023 | Input/output filtering | Filters user prompts and model with an instruction-tuned | ✓ | AUPRC: 84.3 | ToxicChat [212] | N/A |

Table A.8: Overview of deployment-time interventions (part 3).

| Ref. | Year | Method | Description | Model-agnostic | Attack mitigation | Attack | Source code |
|---|---|---|---|---|---|---|---|
| Problem Restatement [80] | 2024 | Self-reflection | Includes instruction to restate the problem in user prompt | ✓ | ASR: 30 → 24 (gpt-3.5-turbo-1106) | Own | N/A |
| [125] | 2024 | Self-reflection | Props model to self-check its outputs using chain of thought (Figure 11) | ✓ | ASR: 49.1 → 0.1 (llama-2-7b-chat) | DAN [25], JADE [226] | N/A |
| NeMo Guardrails [126] | 2023 | Self-reflection | Programmable guardrails | ✓ | ASR: 7 → 1; FP: 2 (gpt-3.5-turbo) | Anthropic's datasets: HH-RLHF [171], [27] | [227] |

Table A.9: Overview of indirect mitigations.

| Ref. | Year | Method | Description | Model-agnostic | Target models | Attack | Source code |
|---|---|---|---|---|---|---|---|
| [45] | 2024 | Watermarking | Uses sequence alignment to embed a distortion-free key sequence into text | ✗ | OPT-1.3B, LLaMA-7B | Own | [228] |
| [46] | 2024 | Watermarking | Trains a generator and detector to inject and detect watermarks | ✗ | OPT-1.3B, llama2-7B | Pegasus [229], DIPPER [230] | [231] |
| KGW [133] | 2023 | Watermarking | Watermarks LLMs' outputs with statistically biased tokens | ✗ | OPT-1.3B | T5-Large [232] | [233] |

Table A.10: Overview of mitigations against indirect prompt injection.

| Ref. | Year | Description | Model-agnostic | Attack mitigation | Attack | Source code |
|---|---|---|---|---|---|---|
| Spotlighting [34] | 2024 | Transforms documents so the model can better distinguish valid from adversarial instructions | ✓ | ASR: 26.3 → 1 (gpt-3.5-turbo) | N/A | N/A |
| IsolateGPT [35] | 2025 | Introduces an isolation system for LLMs through structured agent communication protocols | ✓ | ASR : 20.2 → 7.6 (gpt-4) | InjectAgent [234] | [235] |
| CaMeL [38] | 2025 | Prevents data and control-flow hijacking with a dual-LLM architecture with a safety execution layer | ✓ | ASR: 9.3 → 0 (avg: claude, gemini, gpt, o3, o4) | AgentDojo [157] | [236] |
| SafetyRAG [137] | 2025 | Modular framework that verifies and generates modules to attacks | ✓ | ASR: 65 → 35 (gpt-3.5-turbo), 86 → 7 (llama-2-13b) | Own | N/A |
| Instruction Hierarchy [68] | 2024 | Fine-tuning with synthetic data ro enforce instruction hierarchy: system > user > tool | ✗ | ASR: 37 → 8 (gpt-3.5-turbo) | N/A | N/A |
| ISE [63] | 2024 | Incorporates hierarchy information into embeddings to prioritize instructions | ✗ | ASR: 49.2 → 33.7 (llama) | AlpacaEval [195] | N/A |

Table A.11: Overview of evaluation-time interventions.

| Ref. | Year | Description | Model-agnostic | Source code |
|---|---|---|---|---|
| [141] | 2025 | Evaluates the impact of jailbreak defense strategies on LLMs, measuring their utility, safety, and usability through automated benchmarks | ✓ | N/A |
| PromptShield [139] | 2025 | Benchmarks prompt injection classifiers | ✓ | [237] |
| [140] | 2025 | Quantitatively measures an LLM's ability to separate instructions from data | ✓ | N/A |
| [142] | 2025 | Post-hoc confidence calibration of LLM-based guard models via Temperature Scaling and Contextual Calibration | ✓ | [238] |
| JailbreakBench [143] | 2024 | Benchmarks LLMs for robustness based on a 100 policy-violating behaviors and adversarial prompts | ✓ | [239] |
| Garak [144] | 2024 | A security-probing framework that evaluates LLM vulnerabilities by automatically generating and executing targeted adversarial probes | ✓ | [240] |
| AutoDAN [47] | 2024 | Evolves stealthy jailbreak prompts able to bypass aligned LLM safeguards based on genetic algorithm | ✓ | [241] |
| JailbreakHunter [138] | 2024 | Identifies jailbreak prompts in large-scale human-LLM conversational datasets and provides visual analytics | ✓ | N/A |
| GCG/Advbench [19] | 2023 | Produces adversarial prompts using the gradient-based suffix search method | ✓ | [242] |
| PAIR [26] | 2023 | PAIR utilizes an attacker LLM to continuously refine prompts based on target model responses, achieving effective black-box jailbreaks in fewer than 20 queries | ✓ | [243] |